\documentclass[twocolumn]{revtex4-1}
\usepackage{amsfonts}
\usepackage{amsmath}
\usepackage{xcolor}
\usepackage{graphicx}
\usepackage{hyperref}

\graphicspath{{./}}

\begin{document}

\title{Effects of the surface on double surface Fermi arcs in a realistic tight-binding model of Na$_3$Bi (100).} 
\author{Vasilios K. Passias}
\affiliation{The Anthony J. Leggett Institute for Condensed Matter Theory, and Department of Physics, University of Illinois Urbana-Champaign, Urbana IL 61801, USA} 
\author{Lucas K. Wagner}
\affiliation{The Anthony J. Leggett Institute for Condensed Matter Theory, and Department of Physics, University of Illinois Urbana-Champaign, Urbana IL 61801, USA} 

\begin{abstract}
Na$_3$Bi is a topological Dirac semimetal (TDSM) that can support double surface Fermi arcs (DSFAs), which are important for its material classification and their potential use in spintronic devices. 
However, it is unclear how the surfaces affect realistic Na$_3$Bi (100) systems.
To investigate the effects of diverse surfaces on DSFAs, we consider first principles derived tight-binding models of two Na$_3$Bi (100) terminations.
On the stoichiometric termination, we find DSFAs as expected.
On the non-stoichiometric structure there are DSFAs on one termination and two loops that retain many of the properties of DSFAs.
Thus, the local properties such as spin momentum locking and hybridization with the Dirac points appear to be more robust than their global connectivity.
That is, the surface can reshape the arc fingerprint without destroying the arc-like physics.
\end{abstract}
\maketitle

\section{Introduction}
 
The topological classification of free fermionic materials was expanded to include non-interacting fermionic gapless bulk systems \cite{matsuura_protected_2013,turner_chapter_2013}. 
Such systems include topological semimetals like Weyl and Dirac semimetals \cite{armitage_weyl_2018,gao_topological_2019}.
Here, we focus on Na$_3$Bi, a class I 3D topological Dirac semimetal (TDSM) whose bulk Brillouin zone has two Dirac points at opposite momenta near the origin on the $k_z$ axis  \cite{wang_dirac_2012,yang_classification_2014}, which were observed via ARPES \cite{liu_discovery_2014}.  
Finite size Na$_3$Bi structures are predicted to have double surface Fermi arcs (DSFAs) only on surfaces where the bulk Dirac points are projected onto separate momenta \cite{wang_dirac_2012,potter_quantum_2014,gorbar_surface_2015,gorbar_dirac_2015}.
One such surface is the Na$_3$Bi (100) Miller plane where ARPES measurements showed DSFA features \cite{xu_observation_2015,liang_electronic_2016}.

    \begin{figure*}
        \includegraphics{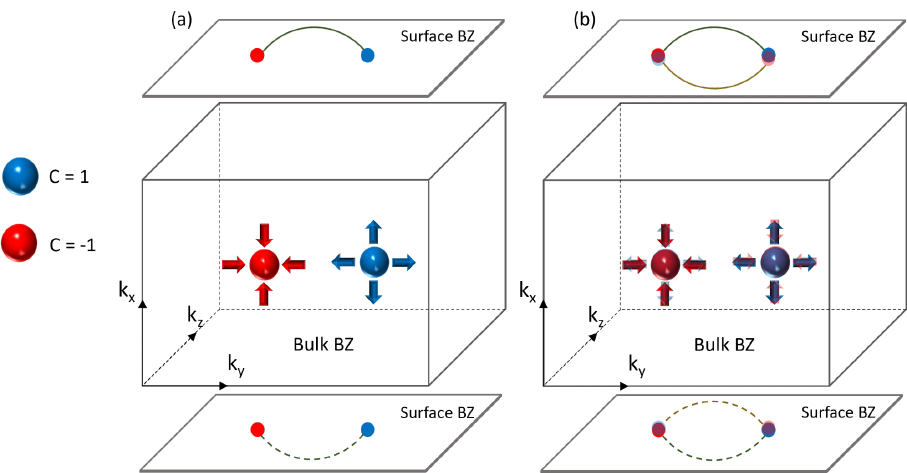}
        \caption{Schematic diagram of Weyl (a) and topological Dirac (b) semimetals with their bulk and surface Brillouin zone features. 
        In (a), we consider a time-reversal symmetry breaking Weyl semimetal with a minimum of two Weyl points having Chern number $C = \pm 1$. 
        A Weyl point with positive (blue) Chern number is a source of Berry curvature field indicated by the outward arrows, whereas inward arrows denote a negative (red) Chern number Weyl point. 
        The dashed Fermi arcs are on the lower surface Brillouin zones. 
        The colored disks on the surface Brillouin zones are the surface projected bulk Weyl points. 
        In (b), the DSFAs on the TDSM terminations are lune shaped and, at the Dirac point energy, are atop one another.
        }
        \label{fig:WSM_DSM_bulk_slab_Fermi_Arcs}
      
    \end{figure*}

Na$_3$Bi's DSFAs are not robustly protected as the surface Fermi arcs in Weyl semimetals \cite{wan_topological_2011,yan_topological_2017}.
Unlike a Weyl semimetal, whose Fermi arcs are surface manifestations of the non-trivial topological invariants of the bulk Weyl points, the DSFAs are not guaranteed from Na$_3$Bi's bulk Dirac points, each of which have zero net Chern number \cite{kargarian_are_2016, kargarian_deformation_2018, wu_fragility_2019,wieder_strong_2020}.
We emphasize that the DSFAs in this paper are not the Fermi arcs between Weyl points of $\pm2$ chiral charge \cite{huang_new_2016} nor the minimum two surface Fermi arcs on a particular termination of time-reversal invariant Weyl semimetals \cite{xu_topological_2018,liu_magnetic_2019,morali_fermi-arc_2019}.
Instead, from the Weyl semimetal Fermi arcs in Fig. \ref{fig:WSM_DSM_bulk_slab_Fermi_Arcs}(a), we consider a schematic of the TDSM DSFAs in Fig. \ref{fig:WSM_DSM_bulk_slab_Fermi_Arcs}(b).

Studies involving four-band tight-binding models of TDSMs suggest that DSFAs are not guaranteed in the presence of bulk perturbations that do not break any bulk symmetries \cite{kargarian_are_2016, kargarian_deformation_2018}. 
Additionally, a low energy effective model of the DSFAs indicated that sufficiently strong surface potentials compared to the Fermi arc curvature potential could remove the DSFAs \cite{potter_quantum_2014}. 
It is unknown if such mechanisms that remove the DSFAs correspond to realistic alterations of pristine Na$_3$Bi systems.
While more realistic calculations were done with density functional theory (DFT) to study charge transfer effects on the Na$_3$Bi $(1\bar{2}0)$ DSFAs, the $(1\bar{2}0)$ systems are limited in thickness from DFT\cite{villanova_engineering_2017}. 
However, there has not been a realistic tight-binding model (based on DFT) that also achieved large enough slabs to resolve the Fermi arcs.

In this manuscript, we investigate whether realistic models of thick finite Na$_3$Bi (100) systems give bulk terms that destroy the Fermi arcs, or if sufficiently perturbed surfaces remove the Fermi arcs.
Tight binding parameters are derived from DFT calculations in contrast to the simplified four-parameter models of Ref~\cite{kargarian_are_2016}. 
We find that for one structure, DSFAs appear as expected, while for a second structure, Fermi arcs appear only on its top termination. 
On its bottom termination there are loops, from the surface and bulk band hybridization, that start and end at the same Dirac points and a larger Fermi loop that encircles the Dirac points.

\section{Background} \label{Background}
Na$_3$Bi's $C_3$ crystal rotation symmetry about the z-axis endows topological invariants to the Dirac points \cite{yang_topological_2015}.
Separately, that same rotation symmetry prevents the Dirac points from being gapped out \cite{yang_classification_2014,wang_dirac_2012} despite each Dirac point being comprised of two Weyl points possessing equal and opposite unit Chern number \cite{armitage_weyl_2018}. 
A related view considered the underlying Weyl points belonging to different and isolated up/down parity sectors thereby classifying Na$_3$Bi as a $\mathbb{Z}_2$ Weyl semimetal\cite{gorbar_surface_2015,gorbar_dirac_2015}.
Hence, both perspectives that have underlying Weyl points indicate that Na$_3$Bi could be viewed as two superimposed Weyl semimetals \cite{potter_quantum_2014,gorbar_surface_2015,gorbar_dirac_2015} like in the bulk of a heuristic TDSM in Fig. \ref{fig:WSM_DSM_bulk_slab_Fermi_Arcs}(b).

From the underlying superimposed Weyl semimetal structure, Na$_3$Bi was originally predicted to have DSFAs, like in Fig.  \ref{fig:WSM_DSM_bulk_slab_Fermi_Arcs}(b).
The individual Fermi arcs connect the momentum separated Dirac points.
We can think of the arcs as being between Weyl points of opposite chiral charge that make up the Dirac points.  
Just like the two coincident Weyl points, the two arcs are prevented from mixing where they meet at the surface projected Dirac points; hence, the lack of mixing gives rise to a discontinuous kink at the surface projected Dirac points \cite{wang_dirac_2012,potter_quantum_2014,kargarian_are_2016}, which we will refer to as a ``cusp.''
These cusps are an important property of the DSFAs that distinguish them from smooth closed Fermi contours.  
This lack of mixing was attributed to the two Weyl points forming each Dirac point belonging to different crystal-rotation irreducible representations, or equivalently, the up and down sectors of Na$_3$Bi's effective Hamiltonian \cite{gorbar_surface_2015,gorbar_dirac_2015}.

Unlike conventional Fermi surfaces, theoretical predictions \cite{potter_quantum_2014,kargarian_are_2016,villanova_engineering_2017} and experiments \cite{xu_observation_2015} verified that the states on the DSFAs are spatially delocalized at and near the surface projected Dirac points, but become localized on slab terminations away from the Dirac point. 
Additionally, the DSFAs were predicted to have vanishing spin magnitude at the projected Dirac points \cite{wang_dirac_2012}.

The DSFAs on the Na$_3$Bi (100) Brillouin zone should intersect the $k_y = 0$ line, which is part of the $k_z = 0$ plane in Na$_3$Bi's bulk Brillouin zone.
That $k_z = 0$ plane has a non-trivial $\mathbb{Z}_2$ invariant, $\nu_{2D} = 1$  \cite{kargarian_are_2016,yang_classification_2014,xu_observation_2015}.
Therefore, this plane has spin-momentum locked, or helical, edge states that are characteristic of the quantum spin Hall effect (QSHE) \cite{ fu_topological_2007,maciejko_quantum_2011}. 
The QSHE on the $k_z = 0$ plane, or $k_y = 0$ on the Na$_3$Bi (100) surface, has two surface localized states per termination with opposite group velocities and spin directions.  
Equivalently, if we consider only half of the Brillouin zone, because of the redundancy of time reversal symmetry, there is only one surface state per termination. 

The properties described above can be summarized as the following criteria, which we use to test for DSFAs in our structures:
\begin{enumerate}
    \item Arcs meet at cusps at surface projected Dirac points \cite{wang_dirac_2012,potter_quantum_2014,kargarian_are_2016}.
    \item At the Dirac point, the states are hybridized between bulk and surface, and localized to the surface away from the Dirac point \cite{potter_quantum_2014, xu_observation_2015, kargarian_are_2016, villanova_engineering_2017}. 
    \item Vanishing spin magnitude at the Dirac points \cite{wang_dirac_2012}.
    \item Odd number of states localized on each surface for $(k_x \geq 0,k_y = 0)$ and constant energy that are singly degenerate with spin-momentum locking \cite{xu_observation_2015,kargarian_are_2016}. 
\end{enumerate}

Properties 1 and 2 are general features of TDSM surface Fermi arcs, also seen in Cd$_3$As$_2$\cite{potter_quantum_2014,kargarian_are_2016}.
Property 1 was not seen in the Na$_3$Bi (1$\bar{2}$0) systems since they were not thick enough given the DFT calculation limitations \cite{villanova_engineering_2017}. 
Property 3 is general to spin-momentum locking and the linear energy-momentum relation near a Dirac point.
Property 4 is associated with the QSHE and is also predicted for Cd$_3$As$_2$'s $k_z = 0$ plane \cite{kargarian_are_2016}.

\section{Method} \label{Method}

Our results were based on a tight-binding (TB) model generated from DFT calculations of the Na$_3$Bi (100) 2-layer periodic structure. 
Supercells were subsequently constructed followed by a modification on the supercell structures to emulate finite sized systems, which we describe in greater detail in the following subsections.

\subsection{Generation of the TB model}

    \begin{figure}
        \includegraphics{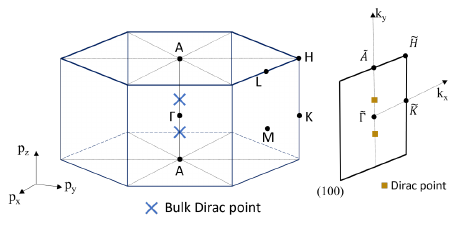}
            \caption{Na$_3$Bi's bulk and (100) surface Brillouin zones with high symmetry k-points and Dirac points  labeled.}
        \label{fig:Na3Bi_bulk_100surf_BZ}
      
    \end{figure}

We used Quantum Espresso (QE) \cite{giannozzi_quantum_2009} to perform DFT calculations on a periodic Na$_3$Bi (100) structure with 2-layers.
We conducted three QE computations on this structure: self-consistent field (scf), followed by non-self consistent field (nscf) and projection (projwfc) calculations. 

In the scf calculation we used a 6 x 4 x 4 Monkhorst-Pack k-space grid \cite{monkhorst_special_1976}, corresponding to a fine spacing of 0.2 \AA$^{-1}$ between k-points, a wavefunction kinetic energy cutoff of 40 Ry and a charge density kinetic energy cutoff of 160 Ry. 
To improve convergence, we used Gaussian smearing with a gaussian spread value of $10^{-3}$ Ry. 
Spin-orbit coupling was implemented by using full relativistic pseudopotentials for both the Bi and Na atoms \cite{dal_corso_bi_nodate, dal_corso_na_nodate}. 
The pseudopotentials used projector augmented waves (PAW) \cite{blochl_projector_1994,kresse_ultrasoft_1999} and the Perdew-Burke-Ernzerhof (PBE) exchange correlation functional type, a subclass of the GGA method \cite{perdew_generalized_1996}.  

The nscf calculation used the same parameters as the scf calculation; however, in the nscf calculation we also specified 260 bands.
Thereafter, the Kohn-Sham eigenstates from the nscf calculations were projected onto an orthonormal basis of Na and Bi atom pseudo-atomic orbitals (PAOs) via the projwfc calculation. 

The results of the three QE  calculations were used by PAOFLOW \cite{buongiorno_nardelli_paoflow_2018,cerasoli_advanced_2021} to construct an accurate TB model for the Na$_3$Bi (100) 2-layer periodic structure.
With four Bi atoms and twelve Na atoms in this structure, the TB model contained $264$ orbitals, or 18 orbitals per Bi atom and 16 orbitals per Na atom.
The parameters of the TB model are included in Appendix \ref{TBvsDFT}, along with a comparison of DFT and TB calculated band structures for the Na$_3$Bi (100) 2-layer periodic structure, which confirm the accuracy of the TB model.

To improve the convergence of the slab calculations, we pinned the Fermi level to the bulk value.
Surface charges could potentially change this Fermi level (particularly in the case of the CU surface); however, we assume these charges are not large enough to materially change the band structure.

\subsection{Generation of slab models from the bulk TB model}

\begin{figure}
    \includegraphics{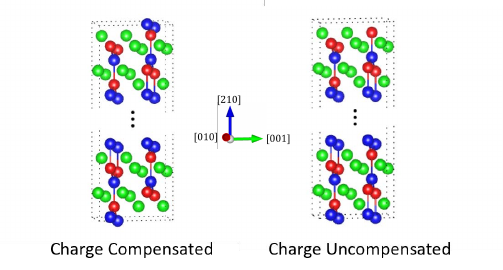}
        \caption{Atomic structure of the two Na$_3$Bi (100) surface terminations considered in this work.  
        The green and blue atoms are the Na atoms, while the red atoms are Bi atoms.  
        The red Bi atoms and blue Na atoms are in the honeycomb layers with bonds between them. 
        The green Na atoms are between the honeycomb layers. 
        We denote [010], [001], and [210] as the $x$-, $y$- and $z$-directions, respectively.}
    \label{fig:CC&CU_structures}
\end{figure}

We used TBmodels \cite{gresch_tbmodels_2024} to create Na$_3$Bi supercell structures from the Na$_3$Bi (100) 2-layer periodic system's baseline TB model. 
We did so by stacking the baseline TB model along the [210] direction, which we denoted as the $z$-direction, perpendicular to the (100) plane; Fig. \ref{fig:Na3Bi_bulk_100surf_BZ} shows the (100) plane. 
Therefore, a (100) supercell consisted of $n_{cells}$ of the Na$_3$Bi (100) 2-layer periodic system, or $2n_{cells}$ layers.
A (100) supercell's Na and Bi PAOs were arranged according to the Na and Bi atom supercell positions, since the PAOs were localized on the atoms. 
Having the PAOs associated with the supercell atoms enabled the creation of two (100) finite sized systems via imposing open boundary conditions along the $z$-direction.

Fig. \ref{fig:CC&CU_structures} shows the two surfaces we considered in this paper. 
One was the charge compensated (CC) system, in which both terminations have three Na atoms and one Bi atom, a stoichiometric arrangement. 
The other structure we term the charge uncompensated (CU) system, because  
unlike CC, CU's terminations had a non-stoichiometric ratio of Na to Bi atoms, with the top termination having one Bi atom and two Na atoms, whereas the bottom termination had only two Na atoms.
Chemically, we believe it is likely that CC is lower energy than CU.

\subsection{Spread from a surface termination}
\label{Spread comp}

Let $\psi_{n,\mathbf{k}_{\parallel},z}$ denote the tight-binding wavefunction of either the CC or CU structures. 
$n$ denotes the band index. 
$\mathbf{k}_{\parallel}$ denotes the crystal momenta, $k_x$, and $k_y$, along the (100) surface Brillouin zone, while $z$ is a layer index along the direction perpendicular to the (100) plane.
The spreads $\sigma_{lower, n, \mathbf{k}_{\parallel} }^2$ and $\sigma_{upper, n, \mathbf{k}_{\parallel}}^2$ are computed with respect to a structure's lower and upper terminations, respectively, as follows:

    \begin{eqnarray}
        \sigma^2_{lower, n, \mathbf{k}_{\parallel}} = \sum _{z = 0}^{L-1} z^{2} | \psi_{n,\mathbf{k}_{\parallel},z}|^2 \label{sprd_lower} \\
        \sigma^2_{upper, n, \mathbf{k}_{\parallel}} = \sum _{z = 0}^{L-1} (L-1 - z)^{2} | \psi_{n,\mathbf{k}_{\parallel},z}|^2 \label{sprd_upper}
    \end{eqnarray}

In equations \ref{sprd_lower} and \ref{sprd_upper}, we sum over the system's discrete number of layers. 
$L$ is the total number of layers.  
Since the number of layers begins from $0$, the sum terminates at $L-1$.  
Therefore, the lower termination is at $z=0$ layers, while the upper termination is at $z=L-1$ layers.
The spread colormap bandplots in Figs. \ref{fig:CC_60layer_var_color_prob_dens} (a) and \ref{fig:CU_60layer_var_color_prob_dens} (a) display the smaller of the two spreads, $\sigma^2_{n, \mathbf{k}_{\parallel}}$, defined as: 

    \begin{equation}
        \sigma^2_{n, \mathbf{k}_{\parallel}} = \text{min}(\sigma^2_{lower, n, \mathbf{k}_{\parallel}},\sigma^2_{upper, n, \mathbf{k}_{\parallel}})    
    \end{equation}

Effectively, at each momentum in the colormap bandplots, $\sigma^2_{n, \mathbf{k}_{\parallel}}$ is the minimum of the two possible spreads.
We use two spreads instead of one, since a localized state with respect to one termination will have a large spread with respect to the opposite termination, and thereby misleadingly indicate a delocalized state. 
Therefore, a small $\sigma^2_{n, \mathbf{k}_{\parallel}}$ indicates a state is localized near a slab termination, but does not specify if the state is localized on the upper or lower termination. 
A large $\sigma^2_{n, \mathbf{k}_{\parallel}}$, however, indicates a state is delocalized.

\subsection{Spin expectation}
\label{Spin-expec-comp}
The spin-expectation value, $\langle \hat{\mathbf{S}} \rangle_{n, \mathbf{k}_{\parallel}}$, is obtained for a single state $n$ at a specific momentum $\mathbf{k}_{\parallel}$ in the (100) surface Brillouin zone. 
$\langle \hat{\mathbf{S}} \rangle_{n, \mathbf{k}_{\parallel}}$ is computed as follows:

    \begin{align}
        \langle \hat{\mathbf{S}} \rangle_{n, \mathbf{k}_{\parallel}} &= \langle \psi_{n,\mathbf{k}_{\parallel}} | \hat{\mathbf{S}} |\psi_{n,\mathbf{k}_{\parallel}} \rangle \notag \\
        & = \sum_{l,j=0}^{N_{orb}-1}\sum_{\chi', \chi} \langle \chi',l| \phi^{*}_{\chi', l,n,\mathbf{k}_{\parallel}}  [\hat{\mathbf{S}}   \otimes \hat{\openone}]   \phi_{\chi,j,n,\mathbf{k}_{\parallel}} |\chi,j\rangle  \notag \\
        &= \sum_{j=0}^{N_{orb}-1}\sum_{\chi', \chi} \phi^{*}_{\chi',j,n,\mathbf{k}_{\parallel}}  \phi_{\chi,j,n,\mathbf{k}_{\parallel}} \langle \chi'|  \hat{\mathbf{S}} |\chi\rangle \label{eqn:spinexpec}
    \end{align}

\noindent Here, $\chi$ and $\chi'$ refer to the spin degrees of freedom while $j$ refers to a Na or Bi orbital (or site). 
$\phi_{\chi,j,n,\mathbf{k}_{\parallel}}$ are the normalized state coefficients.

\subsection{Limitations.}
The TB models for both finite systems are only truncated at the surface and not modified to incorporate surface effects on the model.
We attempted to incorporate surface effects but found that convergence was very slow and we could not perform DFT calculations large enough to obtain accurate surface TB parameters.

Our model also does not include charge effects, so it will miss effects like the Coulomb potential from a surface dipole. 
Nonetheless, we believe these results represent progress towards more realistic models than those considered before\cite{potter_quantum_2014,kargarian_are_2016} because the bulk is modeled very accurately.

\section{Results} \label{Results}

According to the enumerated properties in Section \ref{Background}, we find that the CC structure has DSFAs, which are from both of its terminations and are not separated in momentum space.
The CU structure has Fermi arcs from its top termination only, an outer loop that encloses both surface projected bulk Dirac points, and inner loops that begin and end at the Dirac points.

\subsection{Convergence of Fermi surface with system thickness.}

    \begin{figure}                  
    \includegraphics{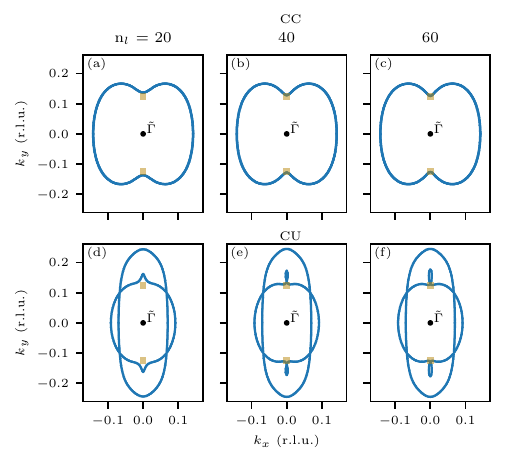}
            \caption{CC's Fermi surface plots, in (a)-(c), for three different thicknesses. 
            n$_l$ denotes the number of layers. 
            The (100) surface projected bulk Dirac points are indicated by dark-goldenrod squares. 
            CU's Fermi surface plots, from (d)-(f), have additional states from its terminations compared to the CC Fermi surfaces. 
            The coordinates are in reciprocal lattice units (r.l.u.)}
        \label{fig:FermiSurf_20_40_60lyrs}
    \end{figure}

Fig.~\ref{fig:FermiSurf_20_40_60lyrs} shows the Fermi surface plots of both structures for three different thicknesses: n$_l = $ 20-, 40-, and 60-layers. 
With increasing thickness, CC's Fermi surface develops cusps at the (100) surface projected bulk Dirac points, which are where the arcs meet as seen in Figs.~\ref{fig:FermiSurf_20_40_60lyrs}(a)-~\ref{fig:FermiSurf_20_40_60lyrs}(c). 
The cusps become sharper with increasing system thickness demonstrating the formation of the Fermi arcs as finite size effects become more negligible and very little change between 40 layers and 60 layers.

In Figs.~\ref{fig:FermiSurf_20_40_60lyrs}(d)-~\ref{fig:FermiSurf_20_40_60lyrs}(f) CU's Fermi surface has an outer loop, which is absent in the CC Fermi surface, that encloses both surface projected Dirac points. 
There are also inner loops that begin and end near the Dirac points, which obscure the cusps, on the CU Fermi surfaces in Figs.~\ref{fig:FermiSurf_20_40_60lyrs}(e)-~\ref{fig:FermiSurf_20_40_60lyrs}(f).
The inner loops become more developed with increasing thickness, changing little between 40 and 60 layers.
Therefore, we focus only on the Fermi surface properties of both 60-layer structures for the remainder of the results.


\subsection{CC DSFA properties.} \label{CC double surface Fermi arcs}

\subsubsection{Cusps of the CC surface Fermi arcs.}

\begin{figure}
    \includegraphics{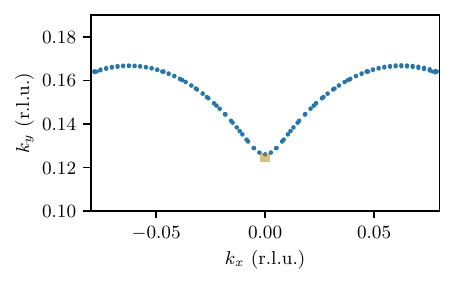}
    \caption{The cusp condition for the CC 60-layer structure's Fermi surface plot, a close-up of Fig.~\ref{fig:FermiSurf_20_40_60lyrs}(c). 
    At the upper Dirac point, denoted by the dark-goldenrod square, a cusp is apparent in the Fermi surface.}
    \label{fig:CC_60layer_slab_Fermi_surf_nearDP}
\end{figure}

CC has surface Fermi arcs that meet at cusps at the surface projected bulk Dirac points, reflecting a standard behavior.
Fig. \ref{fig:CC_60layer_slab_Fermi_surf_nearDP} examines CC's converged Fermi surface plot near one of the (100) surface projected Dirac points.
There we find a cusp where the Fermi arcs meet at the Dirac point, exactly as expected for DSFAs.

    \begin{figure} 
        \centering
        \includegraphics{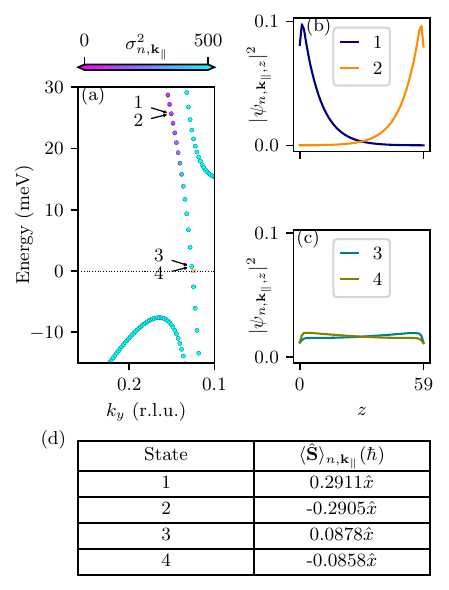}
        
        \caption{CC Fermi arc states spread and suppressed $|\langle \hat{\mathbf{S}} \rangle_{n, \mathbf{k}_{\parallel}}|$ approaching the Dirac point.
        In (a) the colormap bandplot is between $\tilde{A} \rightarrow \tilde{\Gamma}$, along $k_x = 0$ in Fig. \ref{fig:CC_60layer_slab_Fermi_surf_nearDP}. 
        The Dirac point is denoted by the dark-goldenrod square. 
        The colormap indicates the spread ($\sigma^2_{n, \mathbf{k}_{\parallel}}$) of a state with respect to CC's two terminations. 
        Four states are indicated in the colormap and their probability densities are plotted with respect to $z$, perpendicular to the (100) plane, in (b) and (c). In (d) we tabulate
        $\langle \hat{\mathbf{S}} \rangle_{n, \mathbf{k}_{\parallel}}$ of the four states.}
        \label{fig:CC_60layer_var_color_prob_dens}
    \end{figure}

\subsubsection{Hybridization of the bulk and surface and decreasing spin magnitude at the Dirac point.}

In Figs.~\ref{fig:CC_60layer_var_color_prob_dens}(a)-~\ref{fig:CC_60layer_var_color_prob_dens}(c), we show the spread ($\sigma^2_{n, \mathbf{k}_{\parallel}}$) and probability distributions along $z$ of the Fermi arc states along the Fermi arc band. 
States 1 and 2, located away from the Dirac point, are each localized to one of the top and bottom surfaces.
States 3 and 4, very close to the Dirac point, are almost completely delocalized across the entire slab.
Such behavior agrees with the $\sigma^2_{n, \mathbf{k}_{\parallel}}$ colormap in Fig. \ref{fig:CC_60layer_var_color_prob_dens}(a), which shows the Fermi arc states continuously become delocalized (obtain larger $\sigma^2_{n, \mathbf{k}_{\parallel}}$) as they approach the Dirac point.


The spin expectation value magnitude, $|\langle \hat{\mathbf{S}} \rangle_{n, \mathbf{k}_{\parallel}}|$, in Fig.~\ref{fig:CC_60layer_var_color_prob_dens}(d) is about a factor of three smaller for the states closest to the Dirac point, states 3 and 4, than for states 1 and 2 further along the band.
This strong suppression of $|\langle \hat{\mathbf{S}} \rangle_{n, \mathbf{k}_{\parallel}}|$ near the Dirac point is consistent with the vanishing spin magnitude predicted there~\cite{wang_dirac_2012}.
We do not expect the suppression to be exact or strictly monotonic state-by-state: a finite slab cannot host a state exactly at the Dirac point, and the states nearest it are strongly hybridized with the bulk.
We thus conclude that criteria 2 and 3 are satisfied.

    \begin{figure}

        \includegraphics{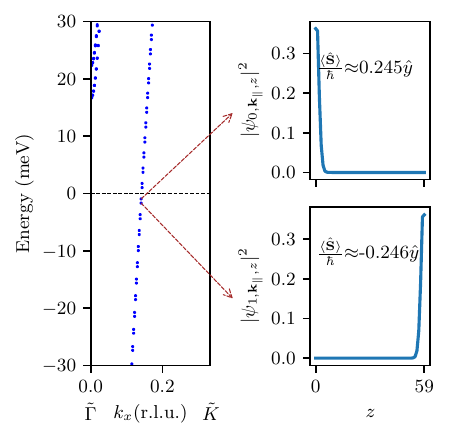}

        \caption{QSHE along $k_y = 0$ intercept of the  CC Fermi arcs.  
        CC's bandplot is along $k_x$, from $\tilde{\Gamma}$ to $\tilde{K}$ in the (100) plane (see Fig. \ref{fig:Na3Bi_bulk_100surf_BZ} for (100) surface Brillouin zone). 
        States on the Fermi arc band nearest the Fermi level have their probability densities and  spin expectation values indicated. 
        Given the presence of time-reversal and inversion symmetries in this structure, the two states have roughly the same  $|\langle \hat{\mathbf{S}} \rangle|$.}
        \label{fig:Charge_compensated_GammaK}

    \end{figure}

\subsubsection{Quantum spin hall effect}

Fig.~\ref{fig:Charge_compensated_GammaK} shows a magnified plot of the CC Fermi arc band intercepted by the $\tilde{\Gamma} \rightarrow \tilde{K}$ line.
There are two bands dispersing with identical positive Fermi velocity in this region.
To demonstrate the quantum spin hall effect (QSHE), we choose two near-degenerate states near the Fermi level, one from each band. 
The state $\psi_{0,\mathbf{k}_\parallel,z}$ localized on the $z=0$ (bottom) of the slab has spin expectation value in the $+\hat{y}$ direction, while the state $\psi_{1,\mathbf{k}_\parallel,z}$ localized on the $z=59$ (top) of the slab has spin expectation value of the same magnitude but exactly the opposite direction, a clear sign of the QSHE.

Based on the properties listed in Section \ref{Background} satisfied by the CC Fermi surface states,  we conclude that the CC Fermi surface is a textbook case of DSFAs.
Since CC is the stoichiometric surface, this conclusion is consistent with previous experiments\cite{xu_observation_2015} that observed DSFAs.
The main difference we find in the numerical models from the DSFAs considered in Fig.~\ref{fig:WSM_DSM_bulk_slab_Fermi_Arcs}(b) is that the CC DSFAs go slightly beyond the Dirac point and then turn inward.
All the topological properties are nonetheless in agreement with the first predictions of Fermi arcs.
Next we will consider the non-stoichiometric CU surface.

\subsection{CU surface Fermi arc properties.} \label{CU surface Fermi arc}

\subsubsection{Cusps of CU surface Fermi arcs.}

    \begin{figure}
        \includegraphics{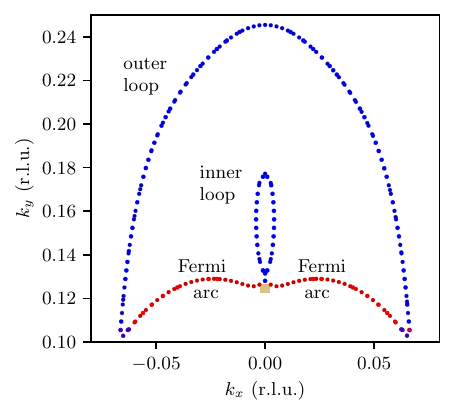}
        \caption{The cusp property of the CU 60-layer structure's Fermi surface, a close-up of Fig.~\ref{fig:FermiSurf_20_40_60lyrs}(f) with a colormap.  
        At the Dirac point, the dark-goldenrod square, there is a cusp. 
        That cusp develops slightly below the inner loop termination. 
        The blue color denotes bottom surface localized states, whereas the red color denotes top surface localized states. 
        } \label{fig:CU_60layer_slab_Fermi_surf_nearDP}
    \end{figure}

As one could have already noted in the finite size study of Fig.~\ref{fig:FermiSurf_20_40_60lyrs}, the CU Fermi surface has three main features, diagrammed in Fig.~\ref{fig:CU_60layer_slab_Fermi_surf_nearDP}. 
These features are the Fermi arc mostly localized to the top surface, which we will show is very similar to the Fermi arcs in the CC surface, and an inner and outer loop.
The inner loop connects to the Dirac point in a cusp, but does not connect to the other Dirac point.
The outer loop does not connect to the Dirac point at all, instead encircling the Dirac points.

\begin{figure}
        \centering
        \includegraphics{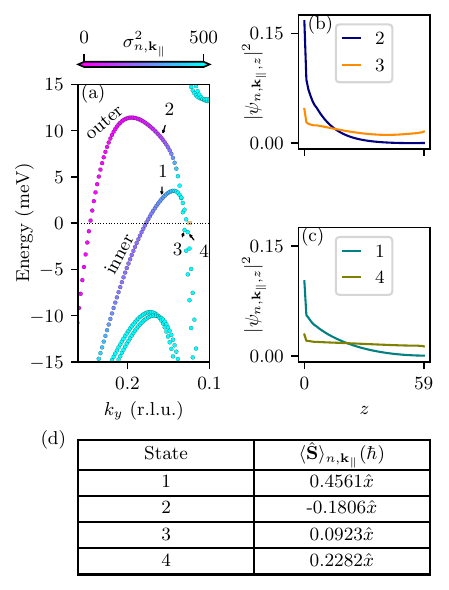}
    
        \caption{CU states' spread and suppressed $|\langle \hat{\mathbf{S}} \rangle_{n, \mathbf{k}_{\parallel}}|$ approaching the Dirac point.
        In (a) the colormap bandplot is between $\tilde{A} \rightarrow \tilde{\Gamma}$, along $k_x = 0$ in Fig.~\ref{fig:CU_60layer_slab_Fermi_surf_nearDP}. 
        ``outer" and ``inner" refer to bands corresponding to the outer and inner loops, respectively.
        The (100) projected bulk Dirac point is denoted by the dark-goldenrod square. 
        The colormap indicates the spread  ($\sigma^2_{n, \mathbf{k}_{\parallel}}$) of a state with respect to its two terminations.   
        Four arbitrarily labeled states away and near the Dirac point are indicated in the colormap with their probability densities plotted with respect to $z$ in (b) and (c). 
        In (d) we tabulate
        $\langle \hat{\mathbf{S}} \rangle_{n, \mathbf{k}_{\parallel}}$ of the four states.
        }
    \label{fig:CU_60layer_var_color_prob_dens}
    \end{figure}

In conclusion, the top and bottom surfaces differ in how their states connect the Dirac points.
On the top surface, a single state runs along the Fermi surface from one Dirac point to the other, as the CC arcs do.
On the bottom surface at the Fermi level no such path exists: the states there form closed loops, an inner loop that touches one Dirac point but not the other, and an outer loop that encircles both.
The outer-loop band does separately reach the Fermi level at the Dirac points in a nearly point-like feature (Fig.~\ref{fig:CU_60layer_var_color_prob_dens}), but this does not create a continuous Fermi-surface path between them.
We interpret these loops as reconstructions of the DSFAs; they retain many of the arcs' properties, but the Dirac points are no longer connected along the Fermi surface.

\subsubsection{Hybridization of the bulk and surface and decreasing spin magnitude at the Dirac point}  

We will focus on the loops here, since the top surface appears to be a classic Fermi arc.
The hybridization and spin properties of the loops are shown in Fig.~\ref{fig:CU_60layer_var_color_prob_dens}. 
Looking at the outer loop (states 2 and 3 in Fig.~\ref{fig:CU_60layer_var_color_prob_dens}(a)), state 2 is localized to the bottom surface (Fig.~\ref{fig:CU_60layer_var_color_prob_dens}(b)), then it crosses the Fermi level again at the Dirac point, which is difficult to see in Fermi surface plots since it only crosses at a very small location.
The outer loop then becomes delocalized at the Dirac point. 
The spin expectation value magnitude drops between states 2 and 3 (Fig.~\ref{fig:CU_60layer_var_color_prob_dens}(d)) as they approach the Dirac point.
This strong suppression is consistent with the vanishing $|\langle \hat{\mathbf{S}} \rangle_{n, \mathbf{k}_{\parallel}}|$ predicted there~\cite{wang_dirac_2012}, though, as for CC, we do not expect it to be exact or strictly monotonic state-by-state, since a finite slab cannot host a state exactly at the Dirac point and the states nearest it are strongly hybridized with the bulk.

The inner loop is apparently more similar to a Fermi arc in the sense that it has a cusp as seen in Fig.~\ref{fig:CU_60layer_slab_Fermi_surf_nearDP}, right above the Fermi arc cusp.
We diagram states 1 and 4 in Fig.~\ref{fig:CU_60layer_var_color_prob_dens}(a) that belong to the inner loop band.
The inner loop band delocalizes as it approaches the Dirac point, and its spin expectation value magnitude (Fig.~\ref{fig:CU_60layer_var_color_prob_dens}(d)) decreases in the same way, as one would expect for a Fermi arc. 

The loops are not trivial surface states: they remain bound to the Dirac points, and the outer loop band, followed away from the Fermi level, connects the two Dirac points even though at the Fermi surface in Fig.~\ref{fig:CU_60layer_slab_Fermi_surf_nearDP} it does not.

\subsubsection{Quantum spin hall effect}
    \begin{figure}
        \includegraphics{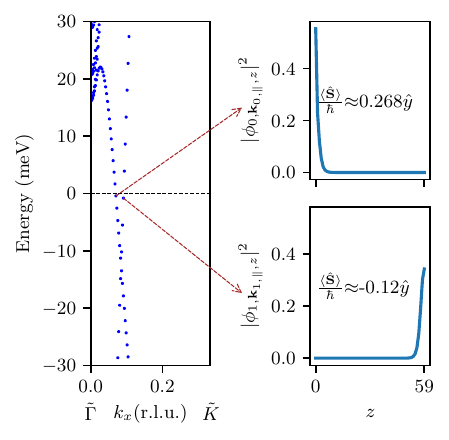}
        \caption{QSHE along $k_y = 0$ segment of the Fermi arcs for CU. 
        The bandplot is along $\tilde{\Gamma} \rightarrow \tilde{K}$ on the (100) plane (see Fig. \ref{fig:Na3Bi_bulk_100surf_BZ} for (100) surface Brillouin zone). 
        States nearest the Fermi level have their probability densities and spin expectation values indicated.
        Because of inversion symmetry breaking it is expected that the two states have distinct $|\langle \hat{\mathbf{S}} \rangle
        |$.
        }
        \label{fig:Charge_uncompensated_GammaK}
    \end{figure}

To check for the presence of a QSHE, we again examined the spin-surface-Fermi velocity relationship along the $\tilde{\Gamma} \rightarrow \tilde{K}$ direction (Fig.~\ref{fig:Charge_uncompensated_GammaK}).
Since the surfaces are not equivalent in the CU geometry, the bands are no longer degenerate with one another. 
The band crossing the Fermi level closer to $\tilde{\Gamma}$ is located on the bottom surface and is thus associated with the ``outer loop'' in Fig.~\ref{fig:CU_60layer_slab_Fermi_surf_nearDP}, while the band crossing it second (further from $\tilde{\Gamma}$) is located on the top surface and thus part of the Fermi arc. 

The two bands on opposite surfaces have Fermi velocities that are in opposite directions, although not the same magnitude.
They have opposing spin polarizations as one would expect for a QSHE. 
However, the spin magnitudes are different by about a factor of two. 
Therefore, this system has a QSHE; however, the different surfaces would have very different transport properties.

\section{Conclusion} \label{Conclusion}

In summary, we constructed a realistic TB model for bulk Na$_3$Bi (100), and produced two finite structures in order to study the effects of the terminations on the surface states.
We found that explicit atomic models of the surface terminations can give rise to complex surface states. 
It's also not an either-or situation; just because there is a Fermi arc on one termination does not mean there is a Fermi arc on the other.
In particular, the CU structure exhibits Fermi arcs on one surface while having two Fermi loops on the other surface.
On the other hand, the stoichiometric CC structure has Fermi arcs on both surfaces.
The CU structure has a QSHE even though it does not have Fermi arcs on the bottom surface, because the QSHE arises from the non-trivial $\mathbb{Z}_2$ invariant of the $k_y=0$ time-reversal invariant plane rather than from the arcs themselves.
The surface can also control the directionality of the Fermi velocity.
In the CC structure, the two states at $k_x > 0$ co-propagate: both are right-moving, with the bottom-surface state carrying $+\hat{y}$ spin and the top-surface state $-\hat{y}$.
In the CU structure the top-surface state is likewise right-moving with $-\hat{y}$ spin, but the bottom-surface state is left-moving while retaining $+\hat{y}$ spin, so the two surfaces counter-propagate and the bottom termination has spin-momentum locking opposite to its CC counterpart.

The surface states of TDSMs are perhaps richer and more varied than previously thought. 
These states are not limited to either forming Fermi arcs or trivial states, but instead loops that hybridize with the Dirac point are possible. 
This could potentially be a lever that allows for control of these states by controlling the surface.




\section*{Acknowledgements}
We acknowledge insightful discussions with Young-Jae Choi, Vatsal Dwivedi, Taylor Hughes, Yoonseok Hwang, W. Joe Meese, William Wheeler, and Benjamin J. Wieder.
We thank Rui Aquino, Young-Jae Choi, Taylor Hughes, and W. Joe Meese for helpful comments on the manuscript.
We thank the Illinois Physics department for generous funding.
We also thank the Illinois Campus Cluster Program for computing time and resources. 
The Illinois Campus Cluster is a computing resource that is operated by the Illinois Campus Cluster Program (ICCP) in conjunction with the National Center for Supercomputing Applications (NCSA) and is supported by funds from the University of Illinois at Urbana-Champaign.

\section*{Data Availability}
The data associated with this paper are available under reasonable request.

\bibliography{ref.bib}

\begin{thebibliography}{37}%
\makeatletter
\providecommand \@ifxundefined [1]{%
 \@ifx{#1\undefined}
}%
\providecommand \@ifnum [1]{%
 \ifnum #1\expandafter \@firstoftwo
 \else \expandafter \@secondoftwo
 \fi
}%
\providecommand \@ifx [1]{%
 \ifx #1\expandafter \@firstoftwo
 \else \expandafter \@secondoftwo
 \fi
}%
\providecommand \natexlab [1]{#1}%
\providecommand \enquote  [1]{``#1''}%
\providecommand \bibnamefont  [1]{#1}%
\providecommand \bibfnamefont [1]{#1}%
\providecommand \citenamefont [1]{#1}%
\providecommand \href@noop [0]{\@secondoftwo}%
\providecommand \href [0]{\begingroup \@sanitize@url \@href}%
\providecommand \@href[1]{\@@startlink{#1}\@@href}%
\providecommand \@@href[1]{\endgroup#1\@@endlink}%
\providecommand \@sanitize@url [0]{\catcode `\\12\catcode `\$12\catcode
  `\&12\catcode `\#12\catcode `\^12\catcode `\_12\catcode `\%12\relax}%
\providecommand \@@startlink[1]{}%
\providecommand \@@endlink[0]{}%
\providecommand \url  [0]{\begingroup\@sanitize@url \@url }%
\providecommand \@url [1]{\endgroup\@href {#1}{\urlprefix }}%
\providecommand \urlprefix  [0]{URL }%
\providecommand \Eprint [0]{\href }%
\providecommand \doibase [0]{http://dx.doi.org/}%
\providecommand \selectlanguage [0]{\@gobble}%
\providecommand \bibinfo  [0]{\@secondoftwo}%
\providecommand \bibfield  [0]{\@secondoftwo}%
\providecommand \translation [1]{[#1]}%
\providecommand \BibitemOpen [0]{}%
\providecommand \bibitemStop [0]{}%
\providecommand \bibitemNoStop [0]{.\EOS\space}%
\providecommand \EOS [0]{\spacefactor3000\relax}%
\providecommand \BibitemShut  [1]{\csname bibitem#1\endcsname}%
\let\auto@bib@innerbib\@empty
\bibitem [{\citenamefont {Matsuura}\ \emph {et~al.}(2013)\citenamefont
  {Matsuura}, \citenamefont {Chang}, \citenamefont {Schnyder},\ and\
  \citenamefont {Ryu}}]{matsuura_protected_2013}%
  \BibitemOpen
  \bibfield  {author} {\bibinfo {author} {\bibfnamefont {S.}~\bibnamefont
  {Matsuura}}, \bibinfo {author} {\bibfnamefont {P.-Y.}\ \bibnamefont {Chang}},
  \bibinfo {author} {\bibfnamefont {A.~P.}\ \bibnamefont {Schnyder}}, \ and\
  \bibinfo {author} {\bibfnamefont {S.}~\bibnamefont {Ryu}},\ }\href {\doibase
  10.1088/1367-2630/15/6/065001} {\bibfield  {journal} {\bibinfo  {journal}
  {New J. Phys.}\ }\textbf {\bibinfo {volume} {15}},\ \bibinfo {pages} {065001}
  (\bibinfo {year} {2013})}\BibitemShut {NoStop}%
\bibitem [{\citenamefont {Turner}\ and\ \citenamefont
  {Vishwanath}(2013)}]{turner_chapter_2013}%
  \BibitemOpen
  \bibfield  {author} {\bibinfo {author} {\bibfnamefont {A.~M.}\ \bibnamefont
  {Turner}}\ and\ \bibinfo {author} {\bibfnamefont {A.}~\bibnamefont
  {Vishwanath}},\ }in\ \href {\doibase 10.1016/B978-0-444-63314-9.00011-1}
  {\emph {\bibinfo {booktitle} {Contemporary {Concepts} of {Condensed} {Matter}
  {Science}}}},\ \bibinfo {series} {Topological {Insulators}}, Vol.~\bibinfo
  {volume} {6},\ \bibinfo {editor} {edited by\ \bibinfo {editor} {\bibfnamefont
  {M.}~\bibnamefont {Franz}}\ and\ \bibinfo {editor} {\bibfnamefont
  {L.}~\bibnamefont {Molenkamp}}}\ (\bibinfo  {publisher} {Elsevier},\ \bibinfo
  {year} {2013})\ pp.\ \bibinfo {pages} {293--324}\BibitemShut {NoStop}%
\bibitem [{\citenamefont {Armitage}\ \emph {et~al.}(2018)\citenamefont
  {Armitage}, \citenamefont {Mele},\ and\ \citenamefont
  {Vishwanath}}]{armitage_weyl_2018}%
  \BibitemOpen
  \bibfield  {author} {\bibinfo {author} {\bibfnamefont {N.}~\bibnamefont
  {Armitage}}, \bibinfo {author} {\bibfnamefont {E.}~\bibnamefont {Mele}}, \
  and\ \bibinfo {author} {\bibfnamefont {A.}~\bibnamefont {Vishwanath}},\
  }\href {\doibase 10.1103/RevModPhys.90.015001} {\bibfield  {journal}
  {\bibinfo  {journal} {Rev. Mod. Phys.}\ }\textbf {\bibinfo {volume} {90}},\
  \bibinfo {pages} {015001} (\bibinfo {year} {2018})}\BibitemShut {NoStop}%
\bibitem [{\citenamefont {Gao}\ \emph {et~al.}(2019)\citenamefont {Gao},
  \citenamefont {Venderbos}, \citenamefont {Kim},\ and\ \citenamefont
  {Rappe}}]{gao_topological_2019}%
  \BibitemOpen
  \bibfield  {author} {\bibinfo {author} {\bibfnamefont {H.}~\bibnamefont
  {Gao}}, \bibinfo {author} {\bibfnamefont {J.~W.}\ \bibnamefont {Venderbos}},
  \bibinfo {author} {\bibfnamefont {Y.}~\bibnamefont {Kim}}, \ and\ \bibinfo
  {author} {\bibfnamefont {A.~M.}\ \bibnamefont {Rappe}},\ }\href {\doibase
  10.1146/annurev-matsci-070218-010049} {\bibfield  {journal} {\bibinfo
  {journal} {Annu. Rev. Mater. Res.}\ }\textbf {\bibinfo {volume} {49}},\
  \bibinfo {pages} {153} (\bibinfo {year} {2019})}\BibitemShut {NoStop}%
\bibitem [{\citenamefont {Wang}\ \emph {et~al.}(2012)\citenamefont {Wang},
  \citenamefont {Sun}, \citenamefont {Chen}, \citenamefont {Franchini},
  \citenamefont {Xu}, \citenamefont {Weng}, \citenamefont {Dai},\ and\
  \citenamefont {Fang}}]{wang_dirac_2012}%
  \BibitemOpen
  \bibfield  {author} {\bibinfo {author} {\bibfnamefont {Z.}~\bibnamefont
  {Wang}}, \bibinfo {author} {\bibfnamefont {Y.}~\bibnamefont {Sun}}, \bibinfo
  {author} {\bibfnamefont {X.-Q.}\ \bibnamefont {Chen}}, \bibinfo {author}
  {\bibfnamefont {C.}~\bibnamefont {Franchini}}, \bibinfo {author}
  {\bibfnamefont {G.}~\bibnamefont {Xu}}, \bibinfo {author} {\bibfnamefont
  {H.}~\bibnamefont {Weng}}, \bibinfo {author} {\bibfnamefont {X.}~\bibnamefont
  {Dai}}, \ and\ \bibinfo {author} {\bibfnamefont {Z.}~\bibnamefont {Fang}},\
  }\href {\doibase 10.1103/PhysRevB.85.195320} {\bibfield  {journal} {\bibinfo
  {journal} {Phys. Rev. B}\ }\textbf {\bibinfo {volume} {85}},\ \bibinfo
  {pages} {195320} (\bibinfo {year} {2012})}\BibitemShut {NoStop}%
\bibitem [{\citenamefont {Yang}\ and\ \citenamefont
  {Nagaosa}(2014)}]{yang_classification_2014}%
  \BibitemOpen
  \bibfield  {author} {\bibinfo {author} {\bibfnamefont {B.-J.}\ \bibnamefont
  {Yang}}\ and\ \bibinfo {author} {\bibfnamefont {N.}~\bibnamefont {Nagaosa}},\
  }\href {\doibase 10.1038/ncomms5898} {\bibfield  {journal} {\bibinfo
  {journal} {Nat Commun}\ }\textbf {\bibinfo {volume} {5}},\ \bibinfo {pages}
  {1} (\bibinfo {year} {2014})},\ \bibinfo {note} {number: 1}\BibitemShut
  {NoStop}%
\bibitem [{\citenamefont {Liu}\ \emph {et~al.}(2014)\citenamefont {Liu},
  \citenamefont {Zhou}, \citenamefont {Zhang}, \citenamefont {Wang},
  \citenamefont {Weng}, \citenamefont {Prabhakaran}, \citenamefont {Mo},
  \citenamefont {Shen}, \citenamefont {Fang}, \citenamefont {Dai},
  \citenamefont {Hussain},\ and\ \citenamefont {Chen}}]{liu_discovery_2014}%
  \BibitemOpen
  \bibfield  {author} {\bibinfo {author} {\bibfnamefont {Z.~K.}\ \bibnamefont
  {Liu}}, \bibinfo {author} {\bibfnamefont {B.}~\bibnamefont {Zhou}}, \bibinfo
  {author} {\bibfnamefont {Y.}~\bibnamefont {Zhang}}, \bibinfo {author}
  {\bibfnamefont {Z.~J.}\ \bibnamefont {Wang}}, \bibinfo {author}
  {\bibfnamefont {H.~M.}\ \bibnamefont {Weng}}, \bibinfo {author}
  {\bibfnamefont {D.}~\bibnamefont {Prabhakaran}}, \bibinfo {author}
  {\bibfnamefont {S.-K.}\ \bibnamefont {Mo}}, \bibinfo {author} {\bibfnamefont
  {Z.~X.}\ \bibnamefont {Shen}}, \bibinfo {author} {\bibfnamefont
  {Z.}~\bibnamefont {Fang}}, \bibinfo {author} {\bibfnamefont {X.}~\bibnamefont
  {Dai}}, \bibinfo {author} {\bibfnamefont {Z.}~\bibnamefont {Hussain}}, \ and\
  \bibinfo {author} {\bibfnamefont {Y.~L.}\ \bibnamefont {Chen}},\ }\href
  {\doibase 10.1126/science.1245085} {\bibfield  {journal} {\bibinfo  {journal}
  {Science}\ }\textbf {\bibinfo {volume} {343}},\ \bibinfo {pages} {864}
  (\bibinfo {year} {2014})}\BibitemShut {NoStop}%
\bibitem [{\citenamefont {Potter}\ \emph {et~al.}(2014)\citenamefont {Potter},
  \citenamefont {Kimchi},\ and\ \citenamefont
  {Vishwanath}}]{potter_quantum_2014}%
  \BibitemOpen
  \bibfield  {author} {\bibinfo {author} {\bibfnamefont {A.~C.}\ \bibnamefont
  {Potter}}, \bibinfo {author} {\bibfnamefont {I.}~\bibnamefont {Kimchi}}, \
  and\ \bibinfo {author} {\bibfnamefont {A.}~\bibnamefont {Vishwanath}},\
  }\href {\doibase 10.1038/ncomms6161} {\bibfield  {journal} {\bibinfo
  {journal} {Nat Commun}\ }\textbf {\bibinfo {volume} {5}},\ \bibinfo {pages}
  {5161} (\bibinfo {year} {2014})}\BibitemShut {NoStop}%
\bibitem [{\citenamefont {Gorbar}\ \emph
  {et~al.}(2015{\natexlab{a}})\citenamefont {Gorbar}, \citenamefont {Miransky},
  \citenamefont {Shovkovy},\ and\ \citenamefont
  {Sukhachov}}]{gorbar_surface_2015}%
  \BibitemOpen
  \bibfield  {author} {\bibinfo {author} {\bibfnamefont {E.~V.}\ \bibnamefont
  {Gorbar}}, \bibinfo {author} {\bibfnamefont {V.~A.}\ \bibnamefont
  {Miransky}}, \bibinfo {author} {\bibfnamefont {I.~A.}\ \bibnamefont
  {Shovkovy}}, \ and\ \bibinfo {author} {\bibfnamefont {P.~O.}\ \bibnamefont
  {Sukhachov}},\ }\href {\doibase 10.1103/PhysRevB.91.235138} {\bibfield
  {journal} {\bibinfo  {journal} {Phys. Rev. B}\ }\textbf {\bibinfo {volume}
  {91}},\ \bibinfo {pages} {235138} (\bibinfo {year}
  {2015}{\natexlab{a}})}\BibitemShut {NoStop}%
\bibitem [{\citenamefont {Gorbar}\ \emph
  {et~al.}(2015{\natexlab{b}})\citenamefont {Gorbar}, \citenamefont {Miransky},
  \citenamefont {Shovkovy},\ and\ \citenamefont
  {Sukhachov}}]{gorbar_dirac_2015}%
  \BibitemOpen
  \bibfield  {author} {\bibinfo {author} {\bibfnamefont {E.~V.}\ \bibnamefont
  {Gorbar}}, \bibinfo {author} {\bibfnamefont {V.~A.}\ \bibnamefont
  {Miransky}}, \bibinfo {author} {\bibfnamefont {I.~A.}\ \bibnamefont
  {Shovkovy}}, \ and\ \bibinfo {author} {\bibfnamefont {P.~O.}\ \bibnamefont
  {Sukhachov}},\ }\href {\doibase 10.1103/PhysRevB.91.121101} {\bibfield
  {journal} {\bibinfo  {journal} {Phys. Rev. B}\ }\textbf {\bibinfo {volume}
  {91}},\ \bibinfo {pages} {121101} (\bibinfo {year}
  {2015}{\natexlab{b}})}\BibitemShut {NoStop}%
\bibitem [{\citenamefont {Xu}\ \emph {et~al.}(2015)\citenamefont {Xu},
  \citenamefont {Liu}, \citenamefont {Kushwaha}, \citenamefont {Sankar},
  \citenamefont {Krizan}, \citenamefont {Belopolski}, \citenamefont {Neupane},
  \citenamefont {Bian}, \citenamefont {Alidoust}, \citenamefont {Chang},
  \citenamefont {Jeng}, \citenamefont {Huang}, \citenamefont {Tsai},
  \citenamefont {Lin}, \citenamefont {Shibayev}, \citenamefont {Chou},
  \citenamefont {Cava},\ and\ \citenamefont {Hasan}}]{xu_observation_2015}%
  \BibitemOpen
  \bibfield  {author} {\bibinfo {author} {\bibfnamefont {S.-Y.}\ \bibnamefont
  {Xu}}, \bibinfo {author} {\bibfnamefont {C.}~\bibnamefont {Liu}}, \bibinfo
  {author} {\bibfnamefont {S.~K.}\ \bibnamefont {Kushwaha}}, \bibinfo {author}
  {\bibfnamefont {R.}~\bibnamefont {Sankar}}, \bibinfo {author} {\bibfnamefont
  {J.~W.}\ \bibnamefont {Krizan}}, \bibinfo {author} {\bibfnamefont
  {I.}~\bibnamefont {Belopolski}}, \bibinfo {author} {\bibfnamefont
  {M.}~\bibnamefont {Neupane}}, \bibinfo {author} {\bibfnamefont
  {G.}~\bibnamefont {Bian}}, \bibinfo {author} {\bibfnamefont {N.}~\bibnamefont
  {Alidoust}}, \bibinfo {author} {\bibfnamefont {T.-R.}\ \bibnamefont {Chang}},
  \bibinfo {author} {\bibfnamefont {H.-T.}\ \bibnamefont {Jeng}}, \bibinfo
  {author} {\bibfnamefont {C.-Y.}\ \bibnamefont {Huang}}, \bibinfo {author}
  {\bibfnamefont {W.-F.}\ \bibnamefont {Tsai}}, \bibinfo {author}
  {\bibfnamefont {H.}~\bibnamefont {Lin}}, \bibinfo {author} {\bibfnamefont
  {P.~P.}\ \bibnamefont {Shibayev}}, \bibinfo {author} {\bibfnamefont {F.-C.}\
  \bibnamefont {Chou}}, \bibinfo {author} {\bibfnamefont {R.~J.}\ \bibnamefont
  {Cava}}, \ and\ \bibinfo {author} {\bibfnamefont {M.~Z.}\ \bibnamefont
  {Hasan}},\ }\href {\doibase 10.1126/science.1256742} {\bibfield  {journal}
  {\bibinfo  {journal} {Science}\ }\textbf {\bibinfo {volume} {347}},\ \bibinfo
  {pages} {294} (\bibinfo {year} {2015})}\BibitemShut {NoStop}%
\bibitem [{\citenamefont {Liang}\ \emph {et~al.}(2016)\citenamefont {Liang},
  \citenamefont {Chen}, \citenamefont {Wang}, \citenamefont {Shi},
  \citenamefont {Feng}, \citenamefont {Yi}, \citenamefont {Xie}, \citenamefont
  {He}, \citenamefont {He}, \citenamefont {Peng}, \citenamefont {Liu},
  \citenamefont {Liu}, \citenamefont {Hu}, \citenamefont {Zhao}, \citenamefont
  {Liu}, \citenamefont {Dong}, \citenamefont {Zhang}, \citenamefont {Nakatake},
  \citenamefont {Iwasawa}, \citenamefont {Shimada}, \citenamefont {Arita},
  \citenamefont {Namatame}, \citenamefont {Taniguchi}, \citenamefont {Xu},
  \citenamefont {Chen}, \citenamefont {Weng}, \citenamefont {Dai},
  \citenamefont {Fang},\ and\ \citenamefont {Zhou}}]{liang_electronic_2016}%
  \BibitemOpen
  \bibfield  {author} {\bibinfo {author} {\bibfnamefont {A.}~\bibnamefont
  {Liang}}, \bibinfo {author} {\bibfnamefont {C.}~\bibnamefont {Chen}},
  \bibinfo {author} {\bibfnamefont {Z.}~\bibnamefont {Wang}}, \bibinfo {author}
  {\bibfnamefont {Y.}~\bibnamefont {Shi}}, \bibinfo {author} {\bibfnamefont
  {Y.}~\bibnamefont {Feng}}, \bibinfo {author} {\bibfnamefont {H.}~\bibnamefont
  {Yi}}, \bibinfo {author} {\bibfnamefont {Z.}~\bibnamefont {Xie}}, \bibinfo
  {author} {\bibfnamefont {S.}~\bibnamefont {He}}, \bibinfo {author}
  {\bibfnamefont {J.}~\bibnamefont {He}}, \bibinfo {author} {\bibfnamefont
  {Y.}~\bibnamefont {Peng}}, \bibinfo {author} {\bibfnamefont {Y.}~\bibnamefont
  {Liu}}, \bibinfo {author} {\bibfnamefont {D.}~\bibnamefont {Liu}}, \bibinfo
  {author} {\bibfnamefont {C.}~\bibnamefont {Hu}}, \bibinfo {author}
  {\bibfnamefont {L.}~\bibnamefont {Zhao}}, \bibinfo {author} {\bibfnamefont
  {G.}~\bibnamefont {Liu}}, \bibinfo {author} {\bibfnamefont {X.}~\bibnamefont
  {Dong}}, \bibinfo {author} {\bibfnamefont {J.}~\bibnamefont {Zhang}},
  \bibinfo {author} {\bibfnamefont {M.}~\bibnamefont {Nakatake}}, \bibinfo
  {author} {\bibfnamefont {H.}~\bibnamefont {Iwasawa}}, \bibinfo {author}
  {\bibfnamefont {K.}~\bibnamefont {Shimada}}, \bibinfo {author} {\bibfnamefont
  {M.}~\bibnamefont {Arita}}, \bibinfo {author} {\bibfnamefont
  {H.}~\bibnamefont {Namatame}}, \bibinfo {author} {\bibfnamefont
  {M.}~\bibnamefont {Taniguchi}}, \bibinfo {author} {\bibfnamefont
  {Z.}~\bibnamefont {Xu}}, \bibinfo {author} {\bibfnamefont {C.}~\bibnamefont
  {Chen}}, \bibinfo {author} {\bibfnamefont {H.}~\bibnamefont {Weng}}, \bibinfo
  {author} {\bibfnamefont {X.}~\bibnamefont {Dai}}, \bibinfo {author}
  {\bibfnamefont {Z.}~\bibnamefont {Fang}}, \ and\ \bibinfo {author}
  {\bibfnamefont {X.-J.}\ \bibnamefont {Zhou}},\ }\href {\doibase
  10.1088/1674-1056/25/7/077101} {\bibfield  {journal} {\bibinfo  {journal}
  {Chinese Phys. B}\ }\textbf {\bibinfo {volume} {25}},\ \bibinfo {pages}
  {077101} (\bibinfo {year} {2016})}\BibitemShut {NoStop}%
\bibitem [{\citenamefont {Wan}\ \emph {et~al.}(2011)\citenamefont {Wan},
  \citenamefont {Turner}, \citenamefont {Vishwanath},\ and\ \citenamefont
  {Savrasov}}]{wan_topological_2011}%
  \BibitemOpen
  \bibfield  {author} {\bibinfo {author} {\bibfnamefont {X.}~\bibnamefont
  {Wan}}, \bibinfo {author} {\bibfnamefont {A.~M.}\ \bibnamefont {Turner}},
  \bibinfo {author} {\bibfnamefont {A.}~\bibnamefont {Vishwanath}}, \ and\
  \bibinfo {author} {\bibfnamefont {S.~Y.}\ \bibnamefont {Savrasov}},\ }\href
  {\doibase 10.1103/PhysRevB.83.205101} {\bibfield  {journal} {\bibinfo
  {journal} {Phys. Rev. B}\ }\textbf {\bibinfo {volume} {83}},\ \bibinfo
  {pages} {205101} (\bibinfo {year} {2011})}\BibitemShut {NoStop}%
\bibitem [{\citenamefont {Yan}\ and\ \citenamefont
  {Felser}(2017)}]{yan_topological_2017}%
  \BibitemOpen
  \bibfield  {author} {\bibinfo {author} {\bibfnamefont {B.}~\bibnamefont
  {Yan}}\ and\ \bibinfo {author} {\bibfnamefont {C.}~\bibnamefont {Felser}},\
  }\href {\doibase 10.1146/annurev-conmatphys-031016-025458} {\bibfield
  {journal} {\bibinfo  {journal} {Annual Review of Condensed Matter Physics}\
  }\textbf {\bibinfo {volume} {8}},\ \bibinfo {pages} {337} (\bibinfo {year}
  {2017})}\BibitemShut {NoStop}%
\bibitem [{\citenamefont {Kargarian}\ \emph {et~al.}(2016)\citenamefont
  {Kargarian}, \citenamefont {Randeria},\ and\ \citenamefont
  {Lu}}]{kargarian_are_2016}%
  \BibitemOpen
  \bibfield  {author} {\bibinfo {author} {\bibfnamefont {M.}~\bibnamefont
  {Kargarian}}, \bibinfo {author} {\bibfnamefont {M.}~\bibnamefont {Randeria}},
  \ and\ \bibinfo {author} {\bibfnamefont {Y.-M.}\ \bibnamefont {Lu}},\ }\href
  {\doibase 10.1073/pnas.1524787113} {\bibfield  {journal} {\bibinfo  {journal}
  {Proc Natl Acad Sci USA}\ }\textbf {\bibinfo {volume} {113}},\ \bibinfo
  {pages} {8648} (\bibinfo {year} {2016})}\BibitemShut {NoStop}%
\bibitem [{\citenamefont {Kargarian}\ \emph {et~al.}(2018)\citenamefont
  {Kargarian}, \citenamefont {Lu},\ and\ \citenamefont
  {Randeria}}]{kargarian_deformation_2018}%
  \BibitemOpen
  \bibfield  {author} {\bibinfo {author} {\bibfnamefont {M.}~\bibnamefont
  {Kargarian}}, \bibinfo {author} {\bibfnamefont {Y.-M.}\ \bibnamefont {Lu}}, \
  and\ \bibinfo {author} {\bibfnamefont {M.}~\bibnamefont {Randeria}},\ }\href
  {\doibase 10.1103/PhysRevB.97.165129} {\bibfield  {journal} {\bibinfo
  {journal} {Phys. Rev. B}\ }\textbf {\bibinfo {volume} {97}},\ \bibinfo
  {pages} {165129} (\bibinfo {year} {2018})}\BibitemShut {NoStop}%
\bibitem [{\citenamefont {Wu}\ \emph {et~al.}(2019)\citenamefont {Wu},
  \citenamefont {Jo}, \citenamefont {Wang}, \citenamefont {Schmidt},
  \citenamefont {Neilson}, \citenamefont {Schrunk}, \citenamefont {Swatek},
  \citenamefont {Eaton}, \citenamefont {Bud'ko}, \citenamefont {Canfield},\
  and\ \citenamefont {Kaminski}}]{wu_fragility_2019}%
  \BibitemOpen
  \bibfield  {author} {\bibinfo {author} {\bibfnamefont {Y.}~\bibnamefont
  {Wu}}, \bibinfo {author} {\bibfnamefont {N.~H.}\ \bibnamefont {Jo}}, \bibinfo
  {author} {\bibfnamefont {L.-L.}\ \bibnamefont {Wang}}, \bibinfo {author}
  {\bibfnamefont {C.~A.}\ \bibnamefont {Schmidt}}, \bibinfo {author}
  {\bibfnamefont {K.~M.}\ \bibnamefont {Neilson}}, \bibinfo {author}
  {\bibfnamefont {B.}~\bibnamefont {Schrunk}}, \bibinfo {author} {\bibfnamefont
  {P.}~\bibnamefont {Swatek}}, \bibinfo {author} {\bibfnamefont
  {A.}~\bibnamefont {Eaton}}, \bibinfo {author} {\bibfnamefont {S.~L.}\
  \bibnamefont {Bud'ko}}, \bibinfo {author} {\bibfnamefont {P.~C.}\
  \bibnamefont {Canfield}}, \ and\ \bibinfo {author} {\bibfnamefont
  {A.}~\bibnamefont {Kaminski}},\ }\href {\doibase 10.1103/PhysRevB.99.161113}
  {\bibfield  {journal} {\bibinfo  {journal} {Phys. Rev. B}\ }\textbf {\bibinfo
  {volume} {99}},\ \bibinfo {pages} {161113} (\bibinfo {year}
  {2019})}\BibitemShut {NoStop}%
\bibitem [{\citenamefont {Wieder}\ \emph {et~al.}(2020)\citenamefont {Wieder},
  \citenamefont {Wang}, \citenamefont {Cano}, \citenamefont {Dai},
  \citenamefont {Schoop}, \citenamefont {Bradlyn},\ and\ \citenamefont
  {Bernevig}}]{wieder_strong_2020}%
  \BibitemOpen
  \bibfield  {author} {\bibinfo {author} {\bibfnamefont {B.~J.}\ \bibnamefont
  {Wieder}}, \bibinfo {author} {\bibfnamefont {Z.}~\bibnamefont {Wang}},
  \bibinfo {author} {\bibfnamefont {J.}~\bibnamefont {Cano}}, \bibinfo {author}
  {\bibfnamefont {X.}~\bibnamefont {Dai}}, \bibinfo {author} {\bibfnamefont
  {L.~M.}\ \bibnamefont {Schoop}}, \bibinfo {author} {\bibfnamefont
  {B.}~\bibnamefont {Bradlyn}}, \ and\ \bibinfo {author} {\bibfnamefont
  {B.~A.}\ \bibnamefont {Bernevig}},\ }\href {\doibase
  10.1038/s41467-020-14443-5} {\bibfield  {journal} {\bibinfo  {journal} {Nat
  Commun}\ }\textbf {\bibinfo {volume} {11}},\ \bibinfo {pages} {627} (\bibinfo
  {year} {2020})},\ \bibinfo {note} {number: 1}\BibitemShut {NoStop}%
\bibitem [{\citenamefont {Huang}\ \emph {et~al.}(2016)\citenamefont {Huang},
  \citenamefont {Xu}, \citenamefont {Belopolski}, \citenamefont {Lee},
  \citenamefont {Chang}, \citenamefont {Chang}, \citenamefont {Wang},
  \citenamefont {Alidoust}, \citenamefont {Bian}, \citenamefont {Neupane},
  \citenamefont {Sanchez}, \citenamefont {Zheng}, \citenamefont {Jeng},
  \citenamefont {Bansil}, \citenamefont {Neupert}, \citenamefont {Lin},\ and\
  \citenamefont {Hasan}}]{huang_new_2016}%
  \BibitemOpen
  \bibfield  {author} {\bibinfo {author} {\bibfnamefont {S.-M.}\ \bibnamefont
  {Huang}}, \bibinfo {author} {\bibfnamefont {S.-Y.}\ \bibnamefont {Xu}},
  \bibinfo {author} {\bibfnamefont {I.}~\bibnamefont {Belopolski}}, \bibinfo
  {author} {\bibfnamefont {C.-C.}\ \bibnamefont {Lee}}, \bibinfo {author}
  {\bibfnamefont {G.}~\bibnamefont {Chang}}, \bibinfo {author} {\bibfnamefont
  {T.-R.}\ \bibnamefont {Chang}}, \bibinfo {author} {\bibfnamefont
  {B.}~\bibnamefont {Wang}}, \bibinfo {author} {\bibfnamefont {N.}~\bibnamefont
  {Alidoust}}, \bibinfo {author} {\bibfnamefont {G.}~\bibnamefont {Bian}},
  \bibinfo {author} {\bibfnamefont {M.}~\bibnamefont {Neupane}}, \bibinfo
  {author} {\bibfnamefont {D.}~\bibnamefont {Sanchez}}, \bibinfo {author}
  {\bibfnamefont {H.}~\bibnamefont {Zheng}}, \bibinfo {author} {\bibfnamefont
  {H.-T.}\ \bibnamefont {Jeng}}, \bibinfo {author} {\bibfnamefont
  {A.}~\bibnamefont {Bansil}}, \bibinfo {author} {\bibfnamefont
  {T.}~\bibnamefont {Neupert}}, \bibinfo {author} {\bibfnamefont
  {H.}~\bibnamefont {Lin}}, \ and\ \bibinfo {author} {\bibfnamefont {M.~Z.}\
  \bibnamefont {Hasan}},\ }\href {\doibase 10.1073/pnas.1514581113} {\bibfield
  {journal} {\bibinfo  {journal} {Proceedings of the National Academy of
  Sciences}\ }\textbf {\bibinfo {volume} {113}},\ \bibinfo {pages} {1180}
  (\bibinfo {year} {2016})}\BibitemShut {NoStop}%
\bibitem [{\citenamefont {Xu}\ \emph {et~al.}(2018)\citenamefont {Xu},
  \citenamefont {Liu}, \citenamefont {Shi}, \citenamefont {Muechler},
  \citenamefont {Gayles}, \citenamefont {Felser},\ and\ \citenamefont
  {Sun}}]{xu_topological_2018}%
  \BibitemOpen
  \bibfield  {author} {\bibinfo {author} {\bibfnamefont {Q.}~\bibnamefont
  {Xu}}, \bibinfo {author} {\bibfnamefont {E.}~\bibnamefont {Liu}}, \bibinfo
  {author} {\bibfnamefont {W.}~\bibnamefont {Shi}}, \bibinfo {author}
  {\bibfnamefont {L.}~\bibnamefont {Muechler}}, \bibinfo {author}
  {\bibfnamefont {J.}~\bibnamefont {Gayles}}, \bibinfo {author} {\bibfnamefont
  {C.}~\bibnamefont {Felser}}, \ and\ \bibinfo {author} {\bibfnamefont
  {Y.}~\bibnamefont {Sun}},\ }\href {\doibase 10.1103/PhysRevB.97.235416}
  {\bibfield  {journal} {\bibinfo  {journal} {Phys. Rev. B}\ }\textbf {\bibinfo
  {volume} {97}},\ \bibinfo {pages} {235416} (\bibinfo {year}
  {2018})}\BibitemShut {NoStop}%
\bibitem [{\citenamefont {Liu}\ \emph {et~al.}(2019)\citenamefont {Liu},
  \citenamefont {Liang}, \citenamefont {Liu}, \citenamefont {Xu}, \citenamefont
  {Li}, \citenamefont {Chen}, \citenamefont {Pei}, \citenamefont {Shi},
  \citenamefont {Mo}, \citenamefont {Dudin}, \citenamefont {Kim}, \citenamefont
  {Cacho}, \citenamefont {Li}, \citenamefont {Sun}, \citenamefont {Yang},
  \citenamefont {Liu}, \citenamefont {Parkin}, \citenamefont {Felser},\ and\
  \citenamefont {Chen}}]{liu_magnetic_2019}%
  \BibitemOpen
  \bibfield  {author} {\bibinfo {author} {\bibfnamefont {D.~F.}\ \bibnamefont
  {Liu}}, \bibinfo {author} {\bibfnamefont {A.~J.}\ \bibnamefont {Liang}},
  \bibinfo {author} {\bibfnamefont {E.~K.}\ \bibnamefont {Liu}}, \bibinfo
  {author} {\bibfnamefont {Q.~N.}\ \bibnamefont {Xu}}, \bibinfo {author}
  {\bibfnamefont {Y.~W.}\ \bibnamefont {Li}}, \bibinfo {author} {\bibfnamefont
  {C.}~\bibnamefont {Chen}}, \bibinfo {author} {\bibfnamefont {D.}~\bibnamefont
  {Pei}}, \bibinfo {author} {\bibfnamefont {W.~J.}\ \bibnamefont {Shi}},
  \bibinfo {author} {\bibfnamefont {S.~K.}\ \bibnamefont {Mo}}, \bibinfo
  {author} {\bibfnamefont {P.}~\bibnamefont {Dudin}}, \bibinfo {author}
  {\bibfnamefont {T.}~\bibnamefont {Kim}}, \bibinfo {author} {\bibfnamefont
  {C.}~\bibnamefont {Cacho}}, \bibinfo {author} {\bibfnamefont
  {G.}~\bibnamefont {Li}}, \bibinfo {author} {\bibfnamefont {Y.}~\bibnamefont
  {Sun}}, \bibinfo {author} {\bibfnamefont {L.~X.}\ \bibnamefont {Yang}},
  \bibinfo {author} {\bibfnamefont {Z.~K.}\ \bibnamefont {Liu}}, \bibinfo
  {author} {\bibfnamefont {S.~S.~P.}\ \bibnamefont {Parkin}}, \bibinfo {author}
  {\bibfnamefont {C.}~\bibnamefont {Felser}}, \ and\ \bibinfo {author}
  {\bibfnamefont {Y.~L.}\ \bibnamefont {Chen}},\ }\href {\doibase
  10.1126/science.aav2873} {\bibfield  {journal} {\bibinfo  {journal}
  {Science}\ }\textbf {\bibinfo {volume} {365}},\ \bibinfo {pages} {1282}
  (\bibinfo {year} {2019})}\BibitemShut {NoStop}%
\bibitem [{\citenamefont {Morali}\ \emph {et~al.}(2019)\citenamefont {Morali},
  \citenamefont {Batabyal}, \citenamefont {Nag}, \citenamefont {Liu},
  \citenamefont {Xu}, \citenamefont {Sun}, \citenamefont {Yan}, \citenamefont
  {Felser}, \citenamefont {Avraham},\ and\ \citenamefont
  {Beidenkopf}}]{morali_fermi-arc_2019}%
  \BibitemOpen
  \bibfield  {author} {\bibinfo {author} {\bibfnamefont {N.}~\bibnamefont
  {Morali}}, \bibinfo {author} {\bibfnamefont {R.}~\bibnamefont {Batabyal}},
  \bibinfo {author} {\bibfnamefont {P.~K.}\ \bibnamefont {Nag}}, \bibinfo
  {author} {\bibfnamefont {E.}~\bibnamefont {Liu}}, \bibinfo {author}
  {\bibfnamefont {Q.}~\bibnamefont {Xu}}, \bibinfo {author} {\bibfnamefont
  {Y.}~\bibnamefont {Sun}}, \bibinfo {author} {\bibfnamefont {B.}~\bibnamefont
  {Yan}}, \bibinfo {author} {\bibfnamefont {C.}~\bibnamefont {Felser}},
  \bibinfo {author} {\bibfnamefont {N.}~\bibnamefont {Avraham}}, \ and\
  \bibinfo {author} {\bibfnamefont {H.}~\bibnamefont {Beidenkopf}},\ }\href
  {\doibase 10.1126/science.aav2334} {\bibfield  {journal} {\bibinfo  {journal}
  {Science}\ }\textbf {\bibinfo {volume} {365}},\ \bibinfo {pages} {1286}
  (\bibinfo {year} {2019})}\BibitemShut {NoStop}%
\bibitem [{\citenamefont {Villanova}\ \emph {et~al.}(2017)\citenamefont
  {Villanova}, \citenamefont {Barnes},\ and\ \citenamefont
  {Park}}]{villanova_engineering_2017}%
  \BibitemOpen
  \bibfield  {author} {\bibinfo {author} {\bibfnamefont {J.~W.}\ \bibnamefont
  {Villanova}}, \bibinfo {author} {\bibfnamefont {E.}~\bibnamefont {Barnes}}, \
  and\ \bibinfo {author} {\bibfnamefont {K.}~\bibnamefont {Park}},\ }\href
  {\doibase 10.1021/acs.nanolett.6b04299} {\bibfield  {journal} {\bibinfo
  {journal} {Nano Lett.}\ }\textbf {\bibinfo {volume} {17}},\ \bibinfo {pages}
  {963} (\bibinfo {year} {2017})}\BibitemShut {NoStop}%
\bibitem [{\citenamefont {Yang}\ \emph {et~al.}(2015)\citenamefont {Yang},
  \citenamefont {Morimoto},\ and\ \citenamefont
  {Furusaki}}]{yang_topological_2015}%
  \BibitemOpen
  \bibfield  {author} {\bibinfo {author} {\bibfnamefont {B.-J.}\ \bibnamefont
  {Yang}}, \bibinfo {author} {\bibfnamefont {T.}~\bibnamefont {Morimoto}}, \
  and\ \bibinfo {author} {\bibfnamefont {A.}~\bibnamefont {Furusaki}},\ }\href
  {\doibase 10.1103/PhysRevB.92.165120} {\bibfield  {journal} {\bibinfo
  {journal} {Phys. Rev. B}\ }\textbf {\bibinfo {volume} {92}},\ \bibinfo
  {pages} {165120} (\bibinfo {year} {2015})}\BibitemShut {NoStop}%
\bibitem [{\citenamefont {Fu}\ and\ \citenamefont
  {Kane}(2007)}]{fu_topological_2007}%
  \BibitemOpen
  \bibfield  {author} {\bibinfo {author} {\bibfnamefont {L.}~\bibnamefont
  {Fu}}\ and\ \bibinfo {author} {\bibfnamefont {C.~L.}\ \bibnamefont {Kane}},\
  }\href {\doibase 10.1103/PhysRevB.76.045302} {\bibfield  {journal} {\bibinfo
  {journal} {Phys. Rev. B}\ }\textbf {\bibinfo {volume} {76}},\ \bibinfo
  {pages} {045302} (\bibinfo {year} {2007})}\BibitemShut {NoStop}%
\bibitem [{\citenamefont {Maciejko}\ \emph {et~al.}(2011)\citenamefont
  {Maciejko}, \citenamefont {Hughes},\ and\ \citenamefont
  {Zhang}}]{maciejko_quantum_2011}%
  \BibitemOpen
  \bibfield  {author} {\bibinfo {author} {\bibfnamefont {J.}~\bibnamefont
  {Maciejko}}, \bibinfo {author} {\bibfnamefont {T.~L.}\ \bibnamefont
  {Hughes}}, \ and\ \bibinfo {author} {\bibfnamefont {S.-C.}\ \bibnamefont
  {Zhang}},\ }\href {\doibase 10.1146/annurev-conmatphys-062910-140538}
  {\bibfield  {journal} {\bibinfo  {journal} {Annu. Rev. Condens. Matter
  Phys.}\ }\textbf {\bibinfo {volume} {2}},\ \bibinfo {pages} {31} (\bibinfo
  {year} {2011})}\BibitemShut {NoStop}%
\bibitem [{\citenamefont {Giannozzi}\ \emph {et~al.}(2009)\citenamefont
  {Giannozzi}, \citenamefont {Baroni}, \citenamefont {Bonini}, \citenamefont
  {Calandra}, \citenamefont {Car}, \citenamefont {Cavazzoni}, \citenamefont
  {Ceresoli}, \citenamefont {Chiarotti}, \citenamefont {Cococcioni},
  \citenamefont {Dabo}, \citenamefont {Corso}, \citenamefont {Gironcoli},
  \citenamefont {Fabris}, \citenamefont {Fratesi}, \citenamefont {Gebauer},
  \citenamefont {Gerstmann}, \citenamefont {Gougoussis}, \citenamefont
  {Kokalj}, \citenamefont {Lazzeri}, \citenamefont {Martin-Samos},
  \citenamefont {Marzari}, \citenamefont {Mauri}, \citenamefont {Mazzarello},
  \citenamefont {Paolini}, \citenamefont {Pasquarello}, \citenamefont
  {Paulatto}, \citenamefont {Sbraccia}, \citenamefont {Scandolo}, \citenamefont
  {Sclauzero}, \citenamefont {Seitsonen}, \citenamefont {Smogunov},
  \citenamefont {Umari},\ and\ \citenamefont
  {Wentzcovitch}}]{giannozzi_quantum_2009}%
  \BibitemOpen
  \bibfield  {author} {\bibinfo {author} {\bibfnamefont {P.}~\bibnamefont
  {Giannozzi}}, \bibinfo {author} {\bibfnamefont {S.}~\bibnamefont {Baroni}},
  \bibinfo {author} {\bibfnamefont {N.}~\bibnamefont {Bonini}}, \bibinfo
  {author} {\bibfnamefont {M.}~\bibnamefont {Calandra}}, \bibinfo {author}
  {\bibfnamefont {R.}~\bibnamefont {Car}}, \bibinfo {author} {\bibfnamefont
  {C.}~\bibnamefont {Cavazzoni}}, \bibinfo {author} {\bibfnamefont
  {D.}~\bibnamefont {Ceresoli}}, \bibinfo {author} {\bibfnamefont {G.~L.}\
  \bibnamefont {Chiarotti}}, \bibinfo {author} {\bibfnamefont {M.}~\bibnamefont
  {Cococcioni}}, \bibinfo {author} {\bibfnamefont {I.}~\bibnamefont {Dabo}},
  \bibinfo {author} {\bibfnamefont {A.~D.}\ \bibnamefont {Corso}}, \bibinfo
  {author} {\bibfnamefont {S.~d.}\ \bibnamefont {Gironcoli}}, \bibinfo {author}
  {\bibfnamefont {S.}~\bibnamefont {Fabris}}, \bibinfo {author} {\bibfnamefont
  {G.}~\bibnamefont {Fratesi}}, \bibinfo {author} {\bibfnamefont
  {R.}~\bibnamefont {Gebauer}}, \bibinfo {author} {\bibfnamefont
  {U.}~\bibnamefont {Gerstmann}}, \bibinfo {author} {\bibfnamefont
  {C.}~\bibnamefont {Gougoussis}}, \bibinfo {author} {\bibfnamefont
  {A.}~\bibnamefont {Kokalj}}, \bibinfo {author} {\bibfnamefont
  {M.}~\bibnamefont {Lazzeri}}, \bibinfo {author} {\bibfnamefont
  {L.}~\bibnamefont {Martin-Samos}}, \bibinfo {author} {\bibfnamefont
  {N.}~\bibnamefont {Marzari}}, \bibinfo {author} {\bibfnamefont
  {F.}~\bibnamefont {Mauri}}, \bibinfo {author} {\bibfnamefont
  {R.}~\bibnamefont {Mazzarello}}, \bibinfo {author} {\bibfnamefont
  {S.}~\bibnamefont {Paolini}}, \bibinfo {author} {\bibfnamefont
  {A.}~\bibnamefont {Pasquarello}}, \bibinfo {author} {\bibfnamefont
  {L.}~\bibnamefont {Paulatto}}, \bibinfo {author} {\bibfnamefont
  {C.}~\bibnamefont {Sbraccia}}, \bibinfo {author} {\bibfnamefont
  {S.}~\bibnamefont {Scandolo}}, \bibinfo {author} {\bibfnamefont
  {G.}~\bibnamefont {Sclauzero}}, \bibinfo {author} {\bibfnamefont {A.~P.}\
  \bibnamefont {Seitsonen}}, \bibinfo {author} {\bibfnamefont {A.}~\bibnamefont
  {Smogunov}}, \bibinfo {author} {\bibfnamefont {P.}~\bibnamefont {Umari}}, \
  and\ \bibinfo {author} {\bibfnamefont {R.~M.}\ \bibnamefont {Wentzcovitch}},\
  }\href {\doibase 10.1088/0953-8984/21/39/395502} {\bibfield  {journal}
  {\bibinfo  {journal} {J. Phys.: Condens. Matter}\ }\textbf {\bibinfo {volume}
  {21}},\ \bibinfo {pages} {395502} (\bibinfo {year} {2009})}\BibitemShut
  {NoStop}%
\bibitem [{\citenamefont {Monkhorst}\ and\ \citenamefont
  {Pack}(1976)}]{monkhorst_special_1976}%
  \BibitemOpen
  \bibfield  {author} {\bibinfo {author} {\bibfnamefont {H.~J.}\ \bibnamefont
  {Monkhorst}}\ and\ \bibinfo {author} {\bibfnamefont {J.~D.}\ \bibnamefont
  {Pack}},\ }\href {\doibase 10.1103/PhysRevB.13.5188} {\bibfield  {journal}
  {\bibinfo  {journal} {Phys. Rev. B}\ }\textbf {\bibinfo {volume} {13}},\
  \bibinfo {pages} {5188} (\bibinfo {year} {1976})}\BibitemShut {NoStop}%
\bibitem [{\citenamefont {Dal~Corso}({\natexlab{a}})}]{dal_corso_bi_nodate}%
  \BibitemOpen
  \bibfield  {author} {\bibinfo {author} {\bibfnamefont {A.}~\bibnamefont
  {Dal~Corso}},\ }\href
  {https://pseudopotentials.quantum-espresso.org/legacy_tables/ps-library/bi}
  {\enquote {\bibinfo {title} {Bi {Full} relativistic {Pseudopotential}.
  https://pseudopotentials.quantum-espresso.org/legacy\_tables/ps-library/bi},}\
  } ({\natexlab{a}})\BibitemShut {NoStop}%
\bibitem [{\citenamefont {Dal~Corso}({\natexlab{b}})}]{dal_corso_na_nodate}%
  \BibitemOpen
  \bibfield  {author} {\bibinfo {author} {\bibfnamefont {A.}~\bibnamefont
  {Dal~Corso}},\ }\href
  {https://pseudopotentials.quantum-espresso.org/legacy_tables/ps-library/na}
  {\enquote {\bibinfo {title} {Na {Full} relativistic {Pseudopotential}.
  https://pseudopotentials.quantum-espresso.org/legacy\_tables/ps-library/na},}\
  } ({\natexlab{b}})\BibitemShut {NoStop}%
\bibitem [{\citenamefont {Blöchl}(1994)}]{blochl_projector_1994}%
  \BibitemOpen
  \bibfield  {author} {\bibinfo {author} {\bibfnamefont {P.~E.}\ \bibnamefont
  {Blöchl}},\ }\href {\doibase 10.1103/PhysRevB.50.17953} {\bibfield
  {journal} {\bibinfo  {journal} {Phys. Rev. B}\ }\textbf {\bibinfo {volume}
  {50}},\ \bibinfo {pages} {17953} (\bibinfo {year} {1994})}\BibitemShut
  {NoStop}%
\bibitem [{\citenamefont {Kresse}\ and\ \citenamefont
  {Joubert}(1999)}]{kresse_ultrasoft_1999}%
  \BibitemOpen
  \bibfield  {author} {\bibinfo {author} {\bibfnamefont {G.}~\bibnamefont
  {Kresse}}\ and\ \bibinfo {author} {\bibfnamefont {D.}~\bibnamefont
  {Joubert}},\ }\href {\doibase 10.1103/PhysRevB.59.1758} {\bibfield  {journal}
  {\bibinfo  {journal} {Phys. Rev. B}\ }\textbf {\bibinfo {volume} {59}},\
  \bibinfo {pages} {1758} (\bibinfo {year} {1999})}\BibitemShut {NoStop}%
\bibitem [{\citenamefont {Perdew}\ \emph {et~al.}(1996)\citenamefont {Perdew},
  \citenamefont {Burke},\ and\ \citenamefont
  {Ernzerhof}}]{perdew_generalized_1996}%
  \BibitemOpen
  \bibfield  {author} {\bibinfo {author} {\bibfnamefont {J.~P.}\ \bibnamefont
  {Perdew}}, \bibinfo {author} {\bibfnamefont {K.}~\bibnamefont {Burke}}, \
  and\ \bibinfo {author} {\bibfnamefont {M.}~\bibnamefont {Ernzerhof}},\ }\href
  {\doibase 10.1103/PhysRevLett.77.3865} {\bibfield  {journal} {\bibinfo
  {journal} {Phys. Rev. Lett.}\ }\textbf {\bibinfo {volume} {77}},\ \bibinfo
  {pages} {3865} (\bibinfo {year} {1996})}\BibitemShut {NoStop}%
\bibitem [{\citenamefont {Buongiorno~Nardelli}\ \emph
  {et~al.}(2018)\citenamefont {Buongiorno~Nardelli}, \citenamefont {Cerasoli},
  \citenamefont {Costa}, \citenamefont {Curtarolo}, \citenamefont {De~Gennaro},
  \citenamefont {Fornari}, \citenamefont {Liyanage}, \citenamefont {Supka},\
  and\ \citenamefont {Wang}}]{buongiorno_nardelli_paoflow_2018}%
  \BibitemOpen
  \bibfield  {author} {\bibinfo {author} {\bibfnamefont {M.}~\bibnamefont
  {Buongiorno~Nardelli}}, \bibinfo {author} {\bibfnamefont {F.~T.}\
  \bibnamefont {Cerasoli}}, \bibinfo {author} {\bibfnamefont {M.}~\bibnamefont
  {Costa}}, \bibinfo {author} {\bibfnamefont {S.}~\bibnamefont {Curtarolo}},
  \bibinfo {author} {\bibfnamefont {R.}~\bibnamefont {De~Gennaro}}, \bibinfo
  {author} {\bibfnamefont {M.}~\bibnamefont {Fornari}}, \bibinfo {author}
  {\bibfnamefont {L.}~\bibnamefont {Liyanage}}, \bibinfo {author}
  {\bibfnamefont {A.~R.}\ \bibnamefont {Supka}}, \ and\ \bibinfo {author}
  {\bibfnamefont {H.}~\bibnamefont {Wang}},\ }\href {\doibase
  10.1016/j.commatsci.2017.11.034} {\bibfield  {journal} {\bibinfo  {journal}
  {Computational Materials Science}\ }\textbf {\bibinfo {volume} {143}},\
  \bibinfo {pages} {462} (\bibinfo {year} {2018})}\BibitemShut {NoStop}%
\bibitem [{\citenamefont {Cerasoli}\ \emph {et~al.}(2021)\citenamefont
  {Cerasoli}, \citenamefont {Supka}, \citenamefont {Jayaraj}, \citenamefont
  {Costa}, \citenamefont {Siloi}, \citenamefont {Sławińska}, \citenamefont
  {Curtarolo}, \citenamefont {Fornari}, \citenamefont {Ceresoli},\ and\
  \citenamefont {Buongiorno~Nardelli}}]{cerasoli_advanced_2021}%
  \BibitemOpen
  \bibfield  {author} {\bibinfo {author} {\bibfnamefont {F.~T.}\ \bibnamefont
  {Cerasoli}}, \bibinfo {author} {\bibfnamefont {A.~R.}\ \bibnamefont {Supka}},
  \bibinfo {author} {\bibfnamefont {A.}~\bibnamefont {Jayaraj}}, \bibinfo
  {author} {\bibfnamefont {M.}~\bibnamefont {Costa}}, \bibinfo {author}
  {\bibfnamefont {I.}~\bibnamefont {Siloi}}, \bibinfo {author} {\bibfnamefont
  {J.}~\bibnamefont {Sławińska}}, \bibinfo {author} {\bibfnamefont
  {S.}~\bibnamefont {Curtarolo}}, \bibinfo {author} {\bibfnamefont
  {M.}~\bibnamefont {Fornari}}, \bibinfo {author} {\bibfnamefont
  {D.}~\bibnamefont {Ceresoli}}, \ and\ \bibinfo {author} {\bibfnamefont
  {M.}~\bibnamefont {Buongiorno~Nardelli}},\ }\href {\doibase
  10.1016/j.commatsci.2021.110828} {\bibfield  {journal} {\bibinfo  {journal}
  {Computational Materials Science}\ }\textbf {\bibinfo {volume} {200}},\
  \bibinfo {pages} {110828} (\bibinfo {year} {2021})}\BibitemShut {NoStop}%
\bibitem [{\citenamefont {Gresch}(2024)}]{gresch_tbmodels_2024}%
  \BibitemOpen
  \bibfield  {author} {\bibinfo {author} {\bibfnamefont {D.}~\bibnamefont
  {Gresch}},\ }\href {https://tbmodels.greschd.ch/en/latest/} {\enquote
  {\bibinfo {title} {{TBmodels} documentation.
  https://tbmodels.greschd.ch/en/latest/},}\ } (\bibinfo {year}
  {2024})\BibitemShut {NoStop}%
\bibitem [{\citenamefont {Di~Bernardo}\ \emph {et~al.}(2020)\citenamefont
  {Di~Bernardo}, \citenamefont {Collins}, \citenamefont {Wu}, \citenamefont
  {Zhou}, \citenamefont {Yang}, \citenamefont {Ju}, \citenamefont {Edmonds},\
  and\ \citenamefont {Fuhrer}}]{di_bernardo_importance_2020}%
  \BibitemOpen
  \bibfield  {author} {\bibinfo {author} {\bibfnamefont {I.}~\bibnamefont
  {Di~Bernardo}}, \bibinfo {author} {\bibfnamefont {J.}~\bibnamefont
  {Collins}}, \bibinfo {author} {\bibfnamefont {W.}~\bibnamefont {Wu}},
  \bibinfo {author} {\bibfnamefont {J.}~\bibnamefont {Zhou}}, \bibinfo {author}
  {\bibfnamefont {S.~A.}\ \bibnamefont {Yang}}, \bibinfo {author}
  {\bibfnamefont {S.}~\bibnamefont {Ju}}, \bibinfo {author} {\bibfnamefont
  {M.~T.}\ \bibnamefont {Edmonds}}, \ and\ \bibinfo {author} {\bibfnamefont
  {M.~S.}\ \bibnamefont {Fuhrer}},\ }\href {\doibase
  10.1103/PhysRevB.102.045124} {\bibfield  {journal} {\bibinfo  {journal}
  {Phys. Rev. B}\ }\textbf {\bibinfo {volume} {102}},\ \bibinfo {pages}
  {045124} (\bibinfo {year} {2020})}\BibitemShut {NoStop}%
\end{thebibliography}%

\newpage
\appendix

\section{Tight-binding model details and validation}

\subsection{PAOFLOW Tight-binding model accuracy.} \label{TBvsDFT}

\begin{figure}
    \includegraphics{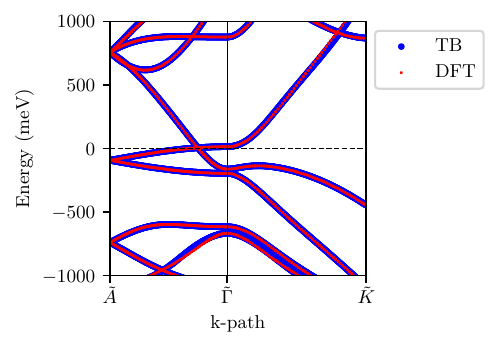}
        \caption{The PAOFLOW tight-binding (TB) model for the Na$_3$Bi (100) 2-layer periodic structure accurately reproduces the density functional theory (DFT) calculations within the energy window of $\pm 1000$ meV about the Fermi level (dashed line).
        } \label{fig:Na3Bi_bulk_100_2lyr_DFTandTB}
\end{figure}

We find that the PAOFLOW pseudo-atomic orbital (PAO) TB model is accurate for our study of Na$_3$Bi (100) slab surface Fermi arcs.
Our appraisal is based on a comparison between DFT and PAOFLOW TB generated bandplots for the Na$_3$Bi (100) 2-layer periodic structure; we evaluate the PAOFLOW TB model accuracy according to its agreement with the DFT bandplot.
The PAO calculations used 400 k-points and the default Gaussian smearing.
The resulting TB and DFT bandplots are presented in Fig. \ref{fig:Na3Bi_bulk_100_2lyr_DFTandTB}.

The bandplot is along the $\tilde{A} \rightarrow \tilde{\Gamma} \rightarrow \tilde{K}$ k-path on the (100) plane. 
The (100) surface Brillouin zone is displayed in Fig \ref{fig:Na3Bi_bulk_100surf_BZ}.
$\tilde{A}, \tilde{\Gamma},$ and $ \tilde{K}$ are some of the high-symmetry k-points of the (100) plane and are represented in PAOFLOW as linear combinations of the (100) plane's reciprocal lattice vectors.
The k-path for this bandplot intercepts one of the (100) projected bulk Dirac points, between $\tilde{A}$ and $\tilde{\Gamma}$ at the Fermi level \cite{wang_dirac_2012}.  
Both the GGA based DFT and TB bands give a tilted Dirac dispersion, which is also observed in the GGA based DFT generated bulk Na$_3$Bi band plot in \cite{wang_dirac_2012} and further reported in \cite{di_bernardo_importance_2020}. 
 
Despite the tilted Dirac dispersion, the TB model and DFT calculations agree over an energy range of $\pm 1$ eV about the Fermi level. 
Beyond this energy window, there are disagreements between the PAOFLOW TB model and DFT calculations. 
However, the agreement over the $\pm 1$ eV energy window is sufficient for our study, since the Fermi arcs appear within a low energy range about the Dirac point. 
Therefore, at this energy range, a scaled up TB model should accurately describe the Fermi arcs on a (100) slab.

\section{Supplementary band-structure and Fermi-surface analysis}

\subsection{CC and CU 60-layer structure bandplot comparisons.} \label{CC&CU_60lyrs_ky_AtoGamma}

    \begin{figure}
        \includegraphics{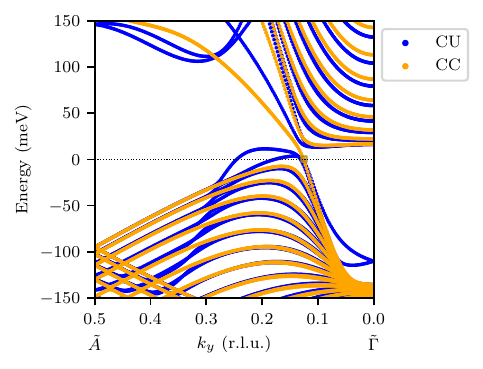} 
        
            \caption{The CC and CU 60-layer band-structures from $\tilde{A} \rightarrow \tilde{\Gamma}$ along $k_y$.  
            The Dirac point is denoted by the dark-goldenrod square.
            These are the expanded energy and momentum range plots in Fig. \ref{fig:CC_60layer_var_color_prob_dens} (a) and Fig. \ref{fig:CU_60layer_var_color_prob_dens} (a) without the colormaps.}
        
        \label{fig:CC&CU_60lyrs}
  
    \end{figure}

Fig. \ref{fig:CC&CU_60lyrs} compares the CC and CU 60-layer bandplots from $\tilde{A}$ to $\tilde{\Gamma}$ along the (100) plane. 
CC has a Fermi arc band descending from the bulk bands above that crosses the Fermi level, near the (100) surface projected bulk Dirac point, and continues to the valence bulk bands below. 

In contrast, CU's terminations break up the energy spectrum flow of the Fermi arc band. 
CU instead has surface bands from approximately 100 meV to 150 meV from $\tilde{A}$ to 0.25 r.l.u., and two bands descending in a similar manner as the CC Fermi arc band, but do not cross the Fermi level.
Additionally, there are two surface bands that form two hole pockets near the Dirac point with the outer loop band forming the longer hole pocket, while inner loop band forms the shorter hole pocket. 
Towards $\tilde{\Gamma}$ the outer and inner loop bands progress below the Fermi level and become degenerate at $\tilde{\Gamma}$.  
The inner and outer loop bands merge into the same band at $k_y \approx 0.295$ r.l.u. and $E\approx-31.3$ meV and continue towards $\tilde{A}$ below the Fermi level. 

The CC and CU bulk bands also have some differences.
Such differences are seen in from approximately -35 meV to -150 meV and from 0.25 r.l.u. to 0.45 r.l.u..
From approximately 0.3 r.l.u. to 0.34 r.l.u. and from -38 meV to -127 meV there are points where the split CU bulk bands merge. 
Also, there are non-overlapping CC and CU bulk bands from 0.3 r.l.u. to 0.5 r.l.u. and energies from -50 meV to -150 meV.
We also find some non-overlapping CC and CU bulk bands above the Fermi level and from approximately 0.15 r.l.u. to $\tilde{\Gamma}$, with the such bands becoming more separate at higher positive energies towards $\tilde{\Gamma}$.

In both systems we find avoided crossings of bands near the Dirac point. 
Such avoided crossings also appear to have a flat dispersion directly above the Dirac point in energy and from approximately 0.1 r.l.u. to $\tilde{\Gamma}$ in momentum space extent.

\subsection{Band behavior of CU outer and inner loops.} \label{CU 60-layer outer and inner}

    \begin{figure*}

        \includegraphics{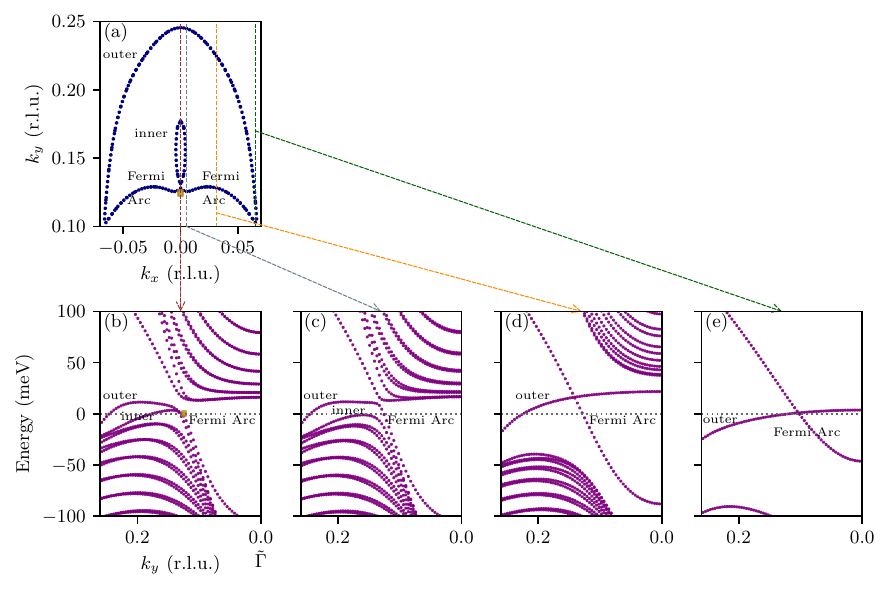}

        \caption{Bandplots along $k_x$ that intercept different CU Fermi surface plot features. 
        The dark-goldenrod square is the (100) surface projected bulk Dirac point.}
        \label{fig:CU_60layer_Fermi_Surf_cuts_bandplot}

    \end{figure*}
    
Fig. \ref{fig:CU_60layer_Fermi_Surf_cuts_bandplot} studies the CU 60-layer system's complex band behavior for regions on its Fermi surface near one of the surface projected bulk Dirac points.
We label three different Fermi surface features in Fig. \ref{fig:CU_60layer_Fermi_Surf_cuts_bandplot} (a): the outer loop, the inner loop, and the Fermi arcs.
We also indicate four different $k_x$ lines along which we obtain four bandplots in Figs. \ref{fig:CU_60layer_Fermi_Surf_cuts_bandplot} (b) - (e).  
All bandplots extend from $k_y = 0.26$ r.l.u. to 0 r.l.u.  
The bandplots should be similar to those near the other surface projected bulk Dirac point for $k_y < 0$.

Along $k_x = 0$ in Fig. \ref{fig:CU_60layer_Fermi_Surf_cuts_bandplot} (b) the outer and inner loop hole pocket bands hybridize with the two bulk bands just above the Fermi level, thereby giving an avoided crossing.
That is contrary to the ``avoidance" of the outer loop states vis-\`{a}-vis the Dirac points as can be inferred from the Fermi surface plot.
At the end of the outer loop hole pocket is the approximate location of the cusp from the top surface Fermi arcs, the cusp being just below the inner loop termination closest to the Dirac point in the Fermi surface. 

The avoided crossing also appears, but with a diminished gap, for the bandplot near $k_x = 0$ in Fig. \ref{fig:CU_60layer_Fermi_Surf_cuts_bandplot} (c).
Additionally, at end of the outer loop hole pocket, the Fermi arc is intercepted.
For larger $k_x$ values in Figs. \ref{fig:CU_60layer_Fermi_Surf_cuts_bandplot} (d) and (e), the bulk bands above and below the Fermi level move further away from the Fermi level.
At the same time, the inner loop is not intercepted as its corresponding band is pushed below the Fermi level. 

In contrast, the lines intercept the Fermi arc and the outer loop.
The corresponding bandplots in Figs. \ref{fig:CU_60layer_Fermi_Surf_cuts_bandplot} (d) and (e) show the Fermi arc band, without spectral interruption, like CC's Fermi arc band in Fig. \ref{fig:CC&CU_60lyrs}, descending from the bulk bands above and passing through the outer loop band without hybridization. 
Instead, there is a gapless crossing between these two bands.
Ultimately, that gapless crossing approaches the Fermi level for larger $k_x$ values, which is seen in the Fig. \ref{fig:CU_60layer_Fermi_Surf_cuts_bandplot} (e) bandplot, which corresponds to a $k_x$ value where the Fermi arc and outer loop intersect in the CU 60-layer's Fermi surface.  
The Fermi arc band then goes through the Fermi level as $k_y$ approaches $0$.  

\subsection{Sampled $k_y$-bandplots from CC 60-layer Fermi surface.} \label{CC 60-layer FS ky-bandplots}

    \begin{figure*}
        \includegraphics{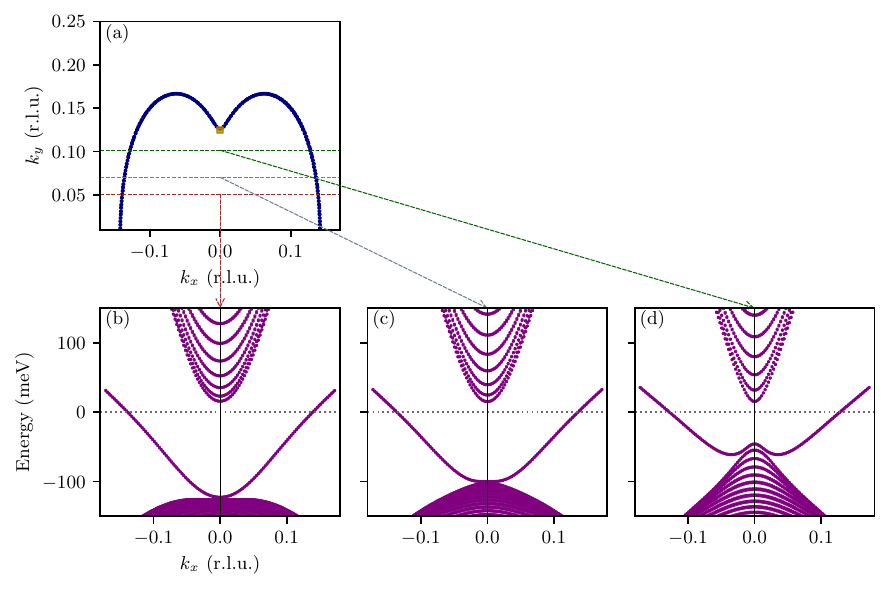}

            \caption{Bandplots along select $k_y$ values from the CC 60-layer Fermi surface data.
            }
        \label{fig:CC_60lyrs_384x256_Ezero_FS_ky_bandplots}
  
    \end{figure*}
From CC's Fermi surface data, we obtain its DSFA band behavior along three selected $k_y$ values in Fig. \ref{fig:CC_60lyrs_384x256_Ezero_FS_ky_bandplots}.
These $k_y$ values are below the dark-goldenrod square that represents the (100) surface projected Dirac point, as seen in Fig. \ref{fig:CC_60lyrs_384x256_Ezero_FS_ky_bandplots}(a).
The corresponding bandplots are in Figs. \ref{fig:CC_60lyrs_384x256_Ezero_FS_ky_bandplots}(b)-(d).

In all of the bandplots, the bulk bands are separated by a bulk band gap, since the sampled $k_y$ values do not intersect the surface projected Dirac point.
In contrast, the DSFA bands pass through the Fermi level at $0$ meV, which agrees with the Fermi surface plot in (a). 
There is no gap separating the DSFA bands from the bulk states in Figs. \ref{fig:CC_60lyrs_384x256_Ezero_FS_ky_bandplots}(b)-(d). 


\subsection{Sampled $k_y$-bandplots from CU 60-layer Fermi surface.}

    \begin{figure*}
        \includegraphics{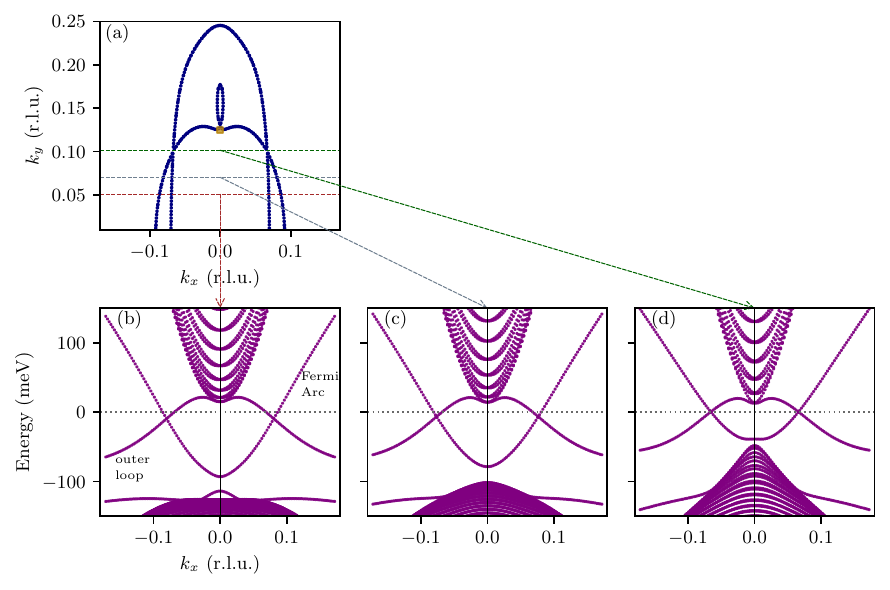}

            \caption{Bandplots along select $k_y$ values from the CU 60-layer Fermi surface data.
            }
        \label{fig:CU_60lyrs_384x256_Ezero_FS_ky_bandplots}
  
    \end{figure*}
Fig. \ref{fig:CU_60lyrs_384x256_Ezero_FS_ky_bandplots} demonstrates the band behavior of the outer loop and Fermi arc bands along three selected $k_y$ values below the Dirac point (dark-goldenrod square in Fig. \ref{fig:CU_60lyrs_384x256_Ezero_FS_ky_bandplots} (a)) from CU's Fermi surface data. 
The corresponding bandplots are in Figs. \ref{fig:CU_60lyrs_384x256_Ezero_FS_ky_bandplots}(b)-(d).
We use the same three $k_y$ values as in Appendix \ref{CC 60-layer FS ky-bandplots}. 

We find a bulk band gap separating the CU bulk states, just like in Fig. \ref{fig:CC_60lyrs_384x256_Ezero_FS_ky_bandplots}. 
Comparably smaller energy gaps separate the bulk bands below the Fermi level from the Fermi arc and outer loop band as seen in Figs. \ref{fig:CU_60lyrs_384x256_Ezero_FS_ky_bandplots} (b) and (c).
That gap decreases as $k_y$ approaches the (100) surface projected bulk Dirac point.
In Fig. \ref{fig:CU_60lyrs_384x256_Ezero_FS_ky_bandplots}(d), there is no gap between the outer loop band and the bulk bands, while there is a small gap between the Fermi arc band and the bulk bands below.

Unlike in Figs. \ref{fig:CC_60lyrs_384x256_Ezero_FS_ky_bandplots} (b)-(d), there is a gap between the Fermi arc band and the bulk states, but only below the Fermi level for the CU 60-layer system. 
In those two figures, the outer loop band will be removed first before the Fermi arc band.
CU's Fermi arcs, established in Section \ref{CU surface Fermi arc}, can be gapped as the chemical potential is lowered relative to the energy of the Dirac points. 
In the context of ARPES measurements on Na$_3$Bi conducted at temperatures equivalent to approximately 0.86 meV to 1.72  meV in \cite{xu_observation_2015}, 2.15 meV in \cite{liang_electronic_2016}, and 8.6 meV in \cite{liu_discovery_2014}, the Fermi arcs should be visible. 
\subsection{CC Fermi surface plots at different Fermi levels.}

    \begin{figure}
        \includegraphics{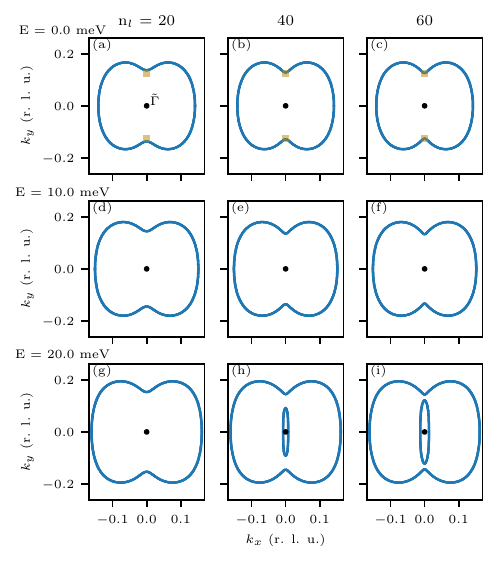} 
        
            \caption{CC Fermi surface plots for different thicknesses and Fermi levels.
            The dark-goldenrod square indicates the (100) surface projected bulk Dirac point.}
        \label{fig:CC_20_40_60_layer_E0.0_E0.01_E0.02_FSplots}
  
    \end{figure}
Fig. \ref{fig:CC_20_40_60_layer_E0.0_E0.01_E0.02_FSplots} presents CC Fermi surface plots at different chemical potentials for thicknesses ranging from 20-layers to 60-layers.
Aside from becoming enlarged, the CC Fermi surfaces in Figs. \ref{fig:CC_20_40_60_layer_E0.0_E0.01_E0.02_FSplots}(d)-(f) do not undergo significant changes as the chemical potential is increased to 10 meV.
At a chemical potential of 20 meV, all CC Fermi surfaces in Figs. \ref{fig:CC_20_40_60_layer_E0.0_E0.01_E0.02_FSplots}(g)-(i) occupy an even greater region of the (100) Brillouin zone. 
However, for 40- and 60-layers in Figs. \ref{fig:CC_20_40_60_layer_E0.0_E0.01_E0.02_FSplots}(h)-(i), an additional Fermi surface feature appears: a loop centered around $\tilde{\Gamma}$.
From 40- to 60-layers, the $\tilde{\Gamma}$ centered loop becomes enlarged.
Unlike the $\tilde{\Gamma}$ centered loop, CC's more persistent DSFA features agrees with the behavior seen in Figs. \ref{fig:CC_60lyrs_384x256_Ezero_FS_ky_bandplots}(b)-(d) for positive chemical potentials.

\subsection{CU Fermi surface plots at different Fermi levels.}

    \begin{figure}
        \includegraphics{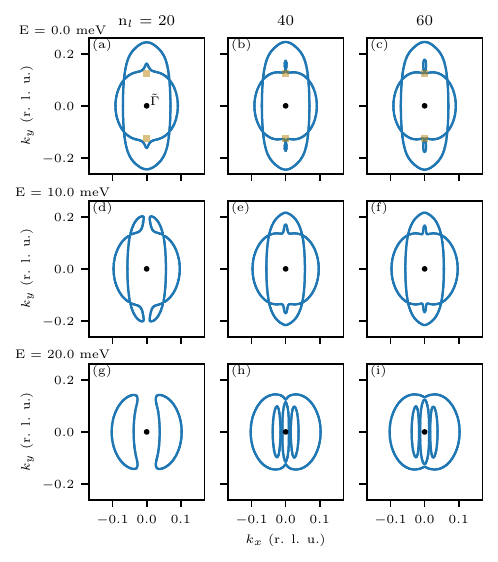} 
        
            \caption{CU Fermi surface plots for different thicknesses and Fermi levels.
            The dark-goldenrod square indicates the (100) surface projected bulk Dirac point.}
        \label{fig:CU_20_40_60_layer_E0.0_E0.01_E0.02_FSplots}
  
    \end{figure}
Fig. \ref{fig:CU_20_40_60_layer_E0.0_E0.01_E0.02_FSplots} has CU Fermi surface plots at the same chemical potentials and thicknesses as in Fig. \ref{fig:CC_60lyrs_384x256_Ezero_FS_ky_bandplots}. 
For 40- and 60-layers, as we increase the chemical potential from 0 meV to 10 meV, the inner loops at the Dirac points and the Fermi arc merge.
At 20 meV, the outer loop and a bulk band give three closely spaced loops near the $\tilde{\Gamma}$ point, for 40- and 60-layers.  
For 40- and 60-layers the loop structure appears to be intact as we increase the chemical potential.
Therefore, while the inner and outer Fermi loops change positions as the chemical potential is increased, they cannot be entirely removed.

One of the three enclosed inner loops is centered about $\tilde{\Gamma}$ and is similar to the $\tilde{\Gamma}$-centered loop in the 40- and 60-layer CC Fermi surface plots at the same chemical potential of 20 meV.
The nested loop centered around $\tilde{\Gamma}$ appears to be from a bulk band; the other two nested loops next to $\tilde{\Gamma}$ appear to be from the splitting of the outer loop. 

The 20-layer structure's Fermi surface plot undergoes the greatest change as we increase the Fermi level. 
From an overall connected Fermi surface plot at $E = 0$ meV, that Fermi surface plot becomes discontinuous at 10 and 20 meV.

\end{document}